\documentclass[fleqn,usenatbib]{rasti}

\usepackage{newtxtext,newtxmath}

\usepackage[T1]{fontenc}

\DeclareRobustCommand{\VAN}[3]{#2}
\let\VANthebibliography\thebibliography
\def\thebibliography{\DeclareRobustCommand{\VAN}[3]{##3}\VANthebibliography}

\usepackage{graphicx}	
\usepackage{amsmath}	
\usepackage{multirow}

\title{Ariel Tier 3 Observations: An Opportunity to Explore both Stellar and Planetary Variability through Broad Wavelength Time-domain Spectroscopy}

\author[A. H. Thompson et al.]{
Alexandra H. Thompson,$^{1,2}$\thanks{E-mail: alex.thompson.17@ucl.ac.uk}
Quentin Changeat,$^{3}$
Theresa Luftinger,$^{4}$
and Giovanna Tinetti$^{5}$
\\
$^{1}$INAF-Osservatorio Astronomico di Palermo, Piazza del Parlamento, 1, 90134 Palermo, Italy\\
$^{2}$Department of Physics and Astronomy, University College London, Gower Street, WC1E 6BT, London, UK\\
$^{3}$Kapteyn Institute, University of Groningen, 9747 AD Groningen, The Netherlands\\
$^{4}$ESA/ESTEC, Science Engagement and Oversight Office (SCI-E), Directorate of Science (SCI)\\
$^{5}$Department of Physics, King’s College London, Strand, London WC2R 2LS, UK\\
}

\date{Accepted XXX. Received YYY; in original form ZZZ}

\pubyear{\the\year{}}

\begin{document}
\label{firstpage}
\pagerange{\pageref{firstpage}--\pageref{lastpage}}
\maketitle

\begin{abstract}
Ariel’s Tier 3 observing strategy offers a unique opportunity to explore planetary and stellar variability in the spectral and time domain. However, without the correct toolkit, temporal variability of the host star may complicate combined analyses of observations obtained at different epochs. We investigate the impact of stellar variability on Ariel Tier 3 observations and establish three different mitigation approaches of increasing physical complexity. Using simulated repeated observations of a favourable system contaminated by different spot and facula configurations, we compare a naive, unmodified stacking approach, separate analysis of the infrared region through progressive removal of the most strongly contaminated optical wavelengths, and the correction of individual visits using combined star-planet retrievals. We find that stacking inconsistent visits can produce substantial biases in the retrieved planetary properties under strong stellar variability, although a single activity correction can still recover the planetary parameters reasonably well. Applying progressive optical data cuts substantially improves the accuracy of retrievals up to a point, without requiring any in-retrieval correction and causing only a modest loss of precision. Correcting individual visits prior to stacking produces more consistent spectra and generally smaller uncertainties, but requires greater computational effort. These results demonstrate that stellar variability should be considered when interpreting repeated Ariel observations, but need not preclude atmospheric characterisation of active systems. A flexible suite of mitigation approaches, supported by preparatory and contemporaneous stellar monitoring, can allow the treatment of stellar variability to be adapted to individual targets while maintaining the efficiency required to maximise Ariel’s scientific return.
\end{abstract}




\section{Introduction}

The exoplanet inventory recently surpassed 6,000 confirmed worlds and is set to grow rapidly over the coming decade, driven by the continued confirmation of TESS candidates \citep{ricker2014}, upcoming Gaia data releases (DR4 and DR5) \citep{gaia2016}, and future missions including ESA's PLATO \citep{rauer2024}, NASA's Nancy Grace Roman Space Telescope \citep{akeson2019}, and CNSA's Earth 2.0 \citep{ge2022}. This expanding population requires new instruments and tools capable of supporting large-scale characterisation.

Scheduled for launch in the early 2030s, ESA's Ariel is a dedicated space-based observatory designed for the population-level characterisation of $\sim$1000 transiting exoplanets over its nominal 4-year mission lifetime. Its instrument suite of three photometers and three spectrometers will provide continuous optical-to-infrared coverage from 0.5–7.8 $\mu$m (Table~\ref{tab:Ariel_specs}), enabling systematic and homogeneous characterisation of star–planet systems on an unprecedented scale \citep{tinetti2018, tinetti2022}. Ariel's design builds on lessons from its predecessors: CoRoT and Kepler demonstrated the importance of repeatable observations over long timescales; HST observations, particularly with the STIS and WFC3 instruments, demonstrated the complementary information encoded across the optical and infrared \citep[e.g.][]{crouzet2014,kreidberg2014,sing2016,lothringer2018,tsiaras2018,pinhas2019,alderson2022,changeat2022,changeat2024,saba2022,saba2025, edwards2023, ramosrosado2025}; and most recently, JWST has highlighted the value of broad wavelength coverage \citep[e.g.][]{carter2024, schlawin2024,welbanks2024, mukherjee2025,kahle2026,liu2026,ma2026,murray2026}, which Ariel leverages by providing optical and infrared observations simultaneously.

Characterising hundreds of planets over four years requires two key ingredients: a target list of observable planets and an efficient scheduling strategy that maximises the scientific return in terms of both planet number and diversity. The Mission Candidate Sample (MCS) provides a continuously updated list of known transiting exoplanets potentially observable with Ariel \citep{zingales2018, edwards2019, edwards2022}, while the Mission Reference Sample (MRS) comprises the $\sim$1000 targets ultimately selected for observation and has yet to be finalised \citep[e.g.][]{edwards2022,cowan2025,panek2026}. The MRS is expected to evolve up to launch as new transiting planets are discovered, for example by TESS and PLATO. Although ideally fixed before launch to enable optimal scheduling, limited changes to the MRS and/or observing schedule may remain possible in exceptional cases where the potential scientific benefit outweighs the associated loss in efficiency (e.g. Kostas Ray et al. in prep.).

\begin{table}
\caption{The wavelength coverages of Ariel's instrument suite and their respective resolutions at each Tier. The line separates the three photometers from the three spectrometers.}
\begin{tabular}{lcccc}
\hline \hline
\multirow{2}{*}{\textbf{Instrument}} & \multirow{2}{*}{\textbf{Wavelength Range ($\mu$m)}} & \multicolumn{3}{c}{\textbf{Resolution}}                                                                \\
                            &                                          & Tier 1                & Tier 2    & Tier 3$^*$  \\ \hline
VISPhot                     & 0.5 - 0.6                                & \multicolumn{3}{c}{\multirow{3}{*}{Photometric bands}}                                        \\
FGS 1                       & 0.6 - 0.8                                & \multicolumn{3}{c}{}                                                                          \\
FGS 2                       & 0.8 - 1.1                                & \multicolumn{3}{c}{}                                                                          \\ \hline
NIRSpec                     & 1.1 - 1.95                               & \multicolumn{1}{l}{$\sim 1$} & $\sim$ 10 & 20                                                        \\
AIRS Ch0                    & 1.95 - 3.9                               & \multicolumn{1}{l}{$\sim 3$} & $\sim$ 50 & 100                                                       \\
AIRS Ch1                    & 3.9 - 7.8                                & \multicolumn{1}{l}{$\sim 1$} & $\sim$ 15 & 30  \\ \hline                                                     
\end{tabular}
\small $^*$Tier 3 uses the native instrument resolution.
\label{tab:Ariel_specs}
\end{table}

\subsection{The Ariel Tier Strategy}

To maximise efficiency and account for the fact that some targets will require greater observing time, Ariel employs a tiered strategy that prioritises targets primarily by their predicted signal-to-noise ratio (SNR), while also maximising planet diversity. The MRS is divided into three main tiers\footnote{A fourth tier, ‘Tier 4’, encompasses phase-curve observations and complementary science cases.} that are characterised by progressively higher spectral resolutions \citep{tinetti2018, edwards2019, mugnai2020}. Targets are categorised according to the number of visits needed to reach each tier's precision requirement, determining the highest tier at which they should feasibly be observed.

Each Tier utilises the full wavelength range of Ariel, albeit with different binning. Tier 1 represents a low-resolution reconnaissance survey comprising three photometric bands and four spectral bands spanning 1.1–7.8 $\mu$m. It aims to exploit the full Ariel sample to address population-level science questions, including the prevalence of primary (H–He dominated) atmospheres, substantial aerosol (cloud/haze) content, and broad trends in atmospheric chemistry \citep{mugnai2021, bocchieri2023, radica2026}. It will also act as a de-risking survey for Tier 2 with the constraints it provides on key system parameters supporting a final optimisation of the Tier 2 target selection. Tier 2 will include approximately half of the targets ($\sim$500 planets) and characterise them at greater depth, with the three photometric bands and the three spectrometers operating at approximately half their native resolutions (Table~\ref{tab:Ariel_specs}). Tier 2 will act as a deep survey, providing more detailed constraints on atmospheric composition, thermal structure, and cloud and haze properties, enabling more robust population-level trends to be identified.

Finally, of greatest relevance to this study, the 50–100 highest-SNR targets, for which an SNR $\geq$7 can be achieved in one or two visits at most, will be observed in Tier 3 at the native instrument resolutions (Table~\ref{tab:Ariel_specs}). The number of visits required depends strongly on system parameters, particularly the magnitude of the host star and planet-to-star radius ratio, $R_{\rm p}/R_*$. This selection effect means that Tier 3 is dominated by comparatively large gaseous planets orbiting bright stars. Repeated visits to these benchmark systems are an integral part of the Tier 3 observing strategy, initially conceived to enable searches of temporal variability of the planetary atmosphere, including weather patterns driven by global atmospheric dynamics \citep[e.g.][]{skinner2022, changeat2024} and variations in heterogeneous cloud coverage \citep[e.g.][]{macdonald2017, cohen2023, loftus2025}. Tier 3 represents one of the unique capabilities of the Ariel mission, the ability to probe system variability with time-domain spectroscopy that would be prohibitively resource-intensive to study with other facilities.

Searching for planetary variability may be especially challenging, signatures of stellar variability can be orders of magnitude larger if the host star is magnetically active. Tier 3 selection is based solely on how observationally favourable the system is i.e. SNR, there is no requirement for a quiet host star. In fact, to fully exploit the scientific potential of Tier 3, systems spanning a range of stellar activity levels should be included. Only through such a sample can we investigate how different timescales of stellar variability affect observations, while also exploring how the changing activity of the host star influences the atmospheric state of planets over time. Additionally, maximising target diversity, consistent with Ariel's broader mission goals, requires representative samples of planets orbiting later-type stars and young systems, both of which are more likely to exhibit strong stellar activity. Smaller, later-type stars provide more favourable planet-to-star radius ratios for characterising smaller planets, while young systems are essential for addressing questions of planet formation and evolution \citep[e.g.][]{turrini2022, pacetti2022, pacetti2025}. Anticipating how this variability may manifest on a visit-to-visit basis is therefore essential to inform the final target selection and begin assembling a toolkit of variability models and mitigation methods prior to Ariel’s launch.

The current MCS contains $\sim$20 planets suitable for Tier 3 transmission observations and $\sim$90 for emission studies. Transit observations are well known to be more susceptible to contributions from stellar activity and variability, however, eclipse studies can also be detrimentally affected by imperfect stellar models \citep{fauchez2025}.
The number of Tier 3 targets may increase as new planets are discovered before Ariel's launch, although most high-SNR systems are likely to have already been identified, with future discoveries preferentially contributing to Tiers 1 and 2. Existing Tier 3 planets include several orbiting renowned active stars such as AU Mic \citep{szabo2021, klein2022, odert2025}, HD 189733 \citep{boisse2009,cauley2017, ehrich2025}, CoRoT-2 \citep{alonso2008,silvavalio2010,bruno2016} and WASP-19 \citep{tregloanreed2013,sedaghati2021,tumborang2024}. Several Tier 3 systems may also be suitable for phase-curve observations \citep{charnay2022}, opening further opportunities to characterise planetary atmospheres in the spatial domain as well as through time.

\subsection{Stellar Variability on Timescales Relevant to the Ariel Mission}

With Ariel's nominal 4-year mission lifetime, and the potential for an extension, Tier 3 targets could be visited repeatedly over several years. This would provide an opportunity to probe variability on timescales ranging from that of a single visit to the full mission duration. Stellar variability is driven by phenomena associated with magnetic activity that vary periodically or quasi-periodically in intensity and/or spatial distribution over time. This variability is consequential for Ariel: sufficiently large changes between visits could render subsequent visits to the same star-planet system inconsistent with one another, requiring individual visits to be treated separately before a robust combined analysis can be performed. The multiplicity of the timescales and their uniqueness to each star also makes variability difficult to predict a priori, highlighting the need for a diverse and flexible toolkit of approaches to identify and mitigate its effects during analysis. In order of increasing characteristic timescales, potentially relevant stellar activity phenomena include oscillations, granulation, flares, photospheric and chromospheric active regions, and activity cycles.

Characterised by the shortest variability timescales i.e. 10s of minutes, oscillation and granulation signals typically have small amplitudes, with Sun-like stars imparting signals on the order of a few ppm in transit observations \citep[e.g.][]{chiavassa2017, sarkar2018, morris2020}. \citet{sarkar2018} performed Ariel-specific sensitivity tests with ExoSim \citep{sarkar2016} and found that the predicted stellar noise from pulsations and granulation falls significantly below the 20 ppm noise floor for all instruments. As such, we reasonably exclude both phenomena from this work.

Stellar flares are impulsive bursts of energy up to $\sim$$10^{36}$ erg with typical durations on the order of hours \citep[e.g.][]{maehara2012}. Superflares substantially increase the incident flux on the exoplanet during and for several hours after the event. This increased flux could modify the atmospheric composition of the planet by driving photochemical reactions and altering the thermal profile \citep{venot2016, louca2023, nicholls2023}. The unpredictability of individual flares means that it is essentially unavoidable that some Ariel observations will be flare contaminated. Contaminated visits are more likely for highly active stars whose flare frequency distributions predict multiple flare events every day \citep{loyd2018, ikuta2023}. This means that flares will undoubtedly be important to consider for Tier 3, especially if planets around bright flare stars e.g. AU Mic b are included in the final MRS. Flare mitigation could be built into the observing strategy by scheduling extra visits to active systems for redundancy, scaled by flare-contamination probability \citep{galletta2026}, however the cost of this would likely be a reduction in the total number of planets observed. At the same time, flare-active systems offer a unique opportunity to use Ariel to probe star–planet interactions and the atmospheric response to transient high-energy stellar irradiation. Their inclusion in Tier 3 would therefore provide a distinct science case, enabling the study of atmospheric variability and disequilibrium chemistry driven by stellar flares alongside Ariel's broader atmospheric survey. This motivates a future dedicated study of Ariel's sensitivity to flare-driven variability, including the atmospheric response to incident flares through disequilibrium chemistry.

The dominant source of stellar variability affecting Ariel is expected to arise from spatially distinct photospheric (spots and faculae) and chromospheric (plage) active regions \citep[e.g.][]{solanki2003,keller2004,berdyugina2005}. The corresponding variability is modulated primarily on the stellar rotation timescale as active regions rotate into and out of view, with feature evolution and decay providing a secondary contribution. However, it is important to note that activity strength does not necessarily translate to variability amplitude: long-lived or continuously replenished regions, such as polar spots, can produce strong but temporally stable contamination \citep[e.g.][]{mori2025}. FGK dwarfs span a wide range of rotation periods, from less than a day to $\sim$100 days. Rotation, age, and activity are closely related, with faster rotators generally being younger and more active because they have undergone less magnetic braking and spin-down \citep[e.g.][]{skumanich1972,pizzolato2003,mamajek2008,gallet2013}. When stellar rotation occurs on timescales comparable to or shorter than the planetary orbital period, even back-to-back observations of successive transits may not fully mitigate the effects of stellar variability.

Importantly, detectable stellar variability is not confined to the most active stars. Kepler observations showed that many stars exhibit measurable photometric variability arising from surface heterogeneities \citep{basri2011,ciardi2011,mcquillan2014}. Indeed, this rotational modulation enabled the measurement of rotation periods for $>$30,000 stars, demonstrating both the prevalence of detectable active regions and the trend of increasing activity towards later spectral types. Evidence for variability and contamination is also emerging in the spectral domain. \citet{saba2025} conducted a population study of 20 high-priority exoplanet systems with archival HST STIS and WFC3 observations, finding widespread evidence for both stellar activity contamination in individual observations and variability between epochs, with more than half of the systems showing potential evidence for some degree of stellar contamination. Recent JWST observations of individual systems similarly suggest that contamination from photospheric active regions may be more pervasive than previously thought \citep[e.g.][]{lim2023,canas2025,espinoza2025,fourniertondreau2025}.

When present, stellar contamination must be accounted for in transmission spectra. If not adequately corrected, the large, highly chromatic signals imparted by stellar heterogeneities can distort or obscure features originating from the planetary atmosphere, significantly biasing their interpretation \citep[e.g.][]{czesla2009,ballerini2012,rackham2018,rackham2019,iyer2020,cracchiolo2021,thompson2024,saba2025}. This is true even for systems in which a single visit is sufficient to achieve the required SNR. For fainter targets requiring multiple visits, however, temporal variability introduces an additional complication, changing configurations of active regions between epochs can render observations of the same system incompatible \citep{espinoza2025,glidden2025,saba2025}. Variable correlated stellar noise can therefore prevent observations from being robustly combined, limiting the expected improvement in SNR from stacking, which scales as $1/\sqrt{N}$ for $N$ independent observations. Multiple visits are a core component of both Tier 2 and Tier 3, although they serve different scientific purposes. Accounting for stellar variability is therefore an important consideration towards realistically assessing Ariel's performance and identifying any mitigation strategies that may be required prior to launch. In this work, we focus on variability arising from changes in the configuration of spots and/or faculae, as these are expected to represent the most prevalent and consequential source of stellar variability for Ariel.

Lastly, Ariel may be sensitive to longer-term activity cycles analogous to the Sun's $\sim$11-year Schwabe cycle \citep{schwabe1844,hathaway2015}. Although Ariel's nominal 4-year mission lifetime is too short to robustly characterise a solar-like activity cycle, it could span some of the shortest cycles detected with periods as short as $\sim$2 years \citep[e.g.][]{borosaikia2018,ayres2025,chahal2025}.

Ariel observations alone are unlikely to be sufficient to confidently identify an activity cycle, but changes in activity level consistent with a star approaching an activity maximum or minimum may nevertheless be detectable. Synergies with PLATO for overlapping targets \citep[e.g.][]{nascimbeni2025}, together with continuous ground-based photometric monitoring and activity indicators such as the \ion{Ca}{ii} H+K lines, could provide complementary constraints on the position of host stars within their activity cycles. In turn, constraints on activity cycles obtained prior to Ariel's launch could inform scheduling decisions, potentially allowing observations to be timed to coincide with stellar activity minima \citep[e.g.][]{niraula2026}. Again, this flexibility would need to be balanced against the necessary efficiency and sample size trade offs.

The importance of accounting for variable stellar activity is heightened by Ariel's population-level science goal. Systematic biases arising from uncorrected stellar contamination or inconsistent treatment across observations could propagate across the sample, distorting the inferred population-level trends that Ariel is designed to measure. The large population Ariel aims to characterise also makes stellar variability particularly challenging to address at scale. Finite resources limit both the extent of preparatory activity monitoring that can be conducted for individual host stars and the extent to which computationally expensive, comprehensive correction models can be applied to every observation. 
In this work, we therefore investigate how changes in stellar activity between repeated Ariel observations can bias the retrieval of planetary atmospheric properties, and assess practical approaches for identifying and mitigating these effects. We compare three methods spanning different levels of physical fidelity and computational cost. By applying these approaches under consistent simulated conditions, we assess their effects on retrieval accuracy and precision and consider how they can provide complementary and efficient strategies for analysing Ariel’s diverse target population. Section~\ref{sec:methodology} introduces the methodology, including the forward modelling of individual contaminated visits and three potential approaches to their subsequent combination and analysis. Section~\ref{sec:results} presents the key results for each approach, with their differences and implications discussed further in Section~\ref{sec:discussion}. Section~\ref{sec:conclusions} concludes with our main findings and recommendations.

\section{Methodology}
\label{sec:methodology}

\subsection{System Choice - HD 209458\,b}

\begin{figure}
    \centering
    \includegraphics[width=\columnwidth]{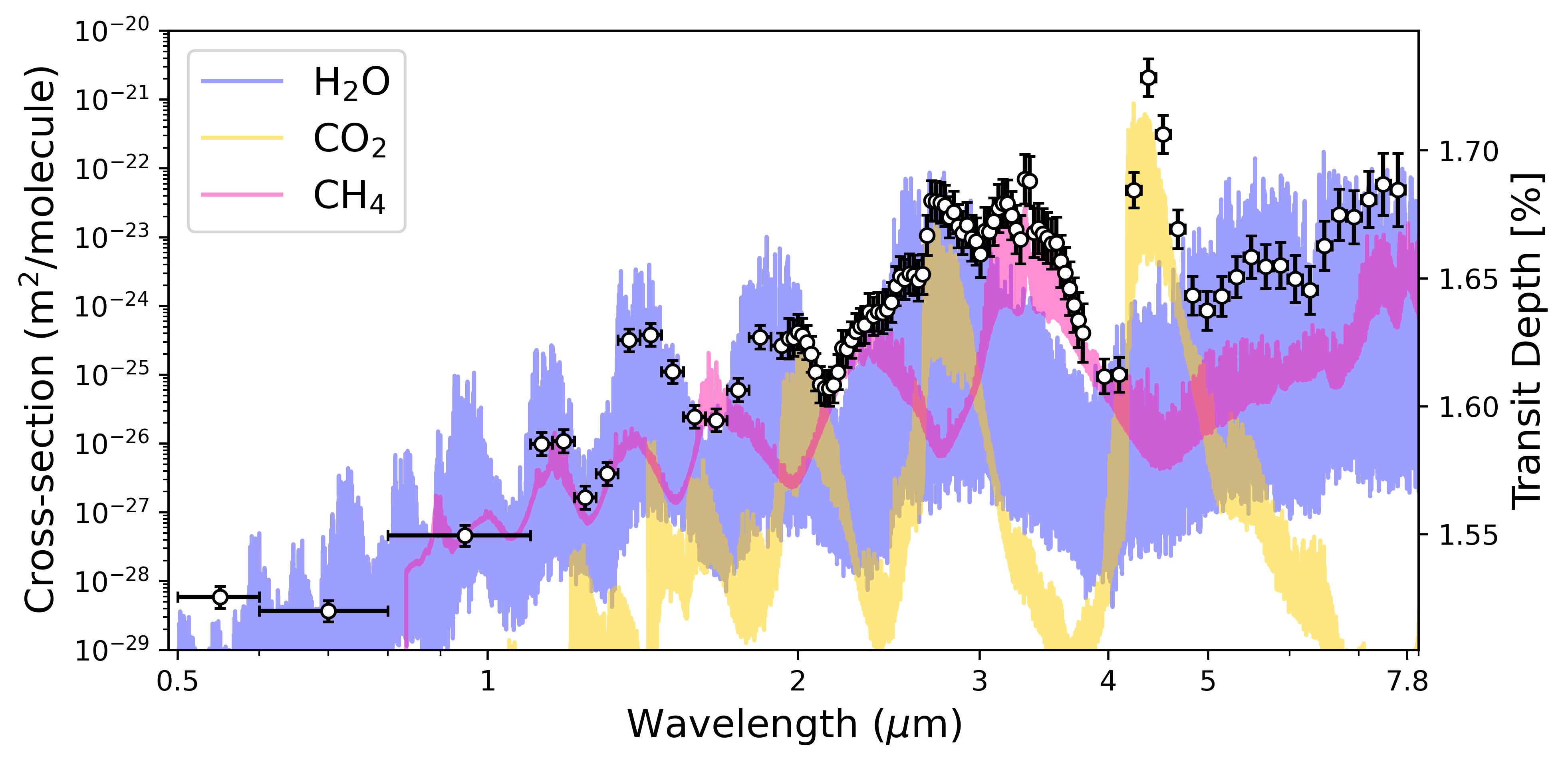}
    \caption{ExoMol cross sections for the active species H$_2$O (\textit{purple}), CO$_2$ (\textit{yellow}), CH$_4$ (\textit{pink}) at a temperature of 1450 K and with mixing ratios of log(H$_2$O)= -3, log(CO$_2$)= -4 and log(CH$_4$) = -4 respectively. The uncontaminated Ariel Tier 3 spectrum is shown by the white data points. Each molecule displays at least one distinct spectral feature at the Tier 3 resolution.}
    \label{fig:cross_secs}
\end{figure}

By design, Tier 3 comprises the most favourable Ariel targets and is therefore biased towards larger planets transiting relatively bright host stars. To assess Ariel's potential sensitivity to stellar variability at Tier 3, we model a system based on HD 209458\,b, one of the highest-SNR targets in the MCS for which a single visit is sufficient to reach Tier 3 resolution. Our aim at this stage is not to perfectly replicate the HD 209458 system, particularly since the star is unlikely to be representative of a highly active or variable host. Its $\log(R^{'}_{\rm HK})$ is consistently measured at $\sim$-4.97, comparable to the Sun during activity minima \citep[e.g.][]{knutson2010,czesla2017,claudi2024}. Instead, the HD 209458 system parameters (Table~\ref{tab:HD209}) provide a convenient baseline for simulating idealised Tier 3 observations, allowing the effects of stellar variability to be isolated more readily. For the same reason, we inject a simplified atmospheric composition at the forward-modelling stage, consisting of three key species (H$_2$O, CO$_2$, and CH$_4$) with high volume mixing ratios. These produce strong absorption features with minimal spectral overlap (Fig.~\ref{fig:cross_secs}), again making any contaminating signals induced by stellar activity and variability easier to disentangle from genuine atmospheric features.

\begin{table}
\caption{Relevant stellar (above line) and planetary parameters (below) for the HD 209458\,b system.}
\centering
\begin{tabular}{lcc}
\hline \hline
\textbf{Parameter}                       & \textbf{Value} \\ \hline
$T_{\rm eff}$ {[}K{]}           & 6091 $\pm$ 10$^{\dagger}$         \\
$R_{*}$ {[}$R_{\odot}${]}       & 1.19 $\pm$ 0.02$^{\dagger}$  \\
$M_{*}$ {[}$M_{\odot}${]}       & 1.23 $\pm$ 0.09$^{\dagger}$  \\
log($g$) {[}cm s$^{-2}${]}      & 4.45 $\pm$ 0.02$^{\dagger}$  \\
{[}Fe/H{]}                      & 0.01$^{\dagger}$ \\
Ks Magnitude                    & 6.308 $\pm$ 0.03  \\ \hline
$T_{\rm eq}$ {[}K{]}            & 1451  \\
$R_{\rm p}$ {[}$R_{\rm Jup}${]} & 1.39 $\pm$ 0.02$^{\dagger}$  \\
$M_{\rm p}$ {[}$M_{\rm Jup}${]} & 0.73 $\pm$ 0.04$^{\dagger}$  \\
log(H$_2$O)                     & -3  \\
log(CO$_2$)                     & -4  \\
log(CH$_4$)                     & -4  \\ \hline
\end{tabular}
\\
\footnotesize{$^\dagger$ \citet{stassun2017}}
\label{tab:HD209}
\end{table}

\subsection{Simulations}

As a first step, we forward model an uncontaminated or nominal transmission spectrum with \texttt{TauREx 3} \citep{alrefaie2021} at the native resolution of the relevant ExoMol linelists \citep{tennyson2024}; H$_2$O \citep{polyansky2018}, CO$_2$ \citep{yurchenko2020} and CH$_4$ \citep{yurchenko2024}. We assume a primary H and He dominated atmosphere with a ratio of 0.17 with the three active species present as trace gases.
We assume an isothermal temperature profile fixed to the value given in the Ariel MCS to ensure consistency with other Ariel-specific studies and use a standard pressure profile spanning 1 - 10$^6$ Pa with 100 layers. Contributions from Rayleigh scattering and collision-induced absorption (CIA) from H-H and H-He pairs are also included. This uncontaminated spectrum is subsequently convolved with a grid of wavelength-dependent contamination spectra generated with the stellar activity mixin ASteRA \citep{thompson2024} before finally being binned to the Ariel Tier 3 resolution and assigned wavelength-dependent uncertainties estimated with ArielRad \citep{mugnai2020}. 

To explore the potential range of stellar-variability effects, we generated a grid of 30 spectra contaminated by either spots or faculae, spanning a range of temperature contrasts and filling factors motivated by observed heterogeneities and therefore representative of those that Ariel may encounter. Contaminated spectra are produced through the weighted combination of multiple stellar spectral energy distributions (SEDs) from the PHOENIX BT-Settl library \citep{allard2012}, for more information about ASteRA see \citet{thompson2024}. For the 15 spot-contaminated spectra, we estimated the spot temperature using the updated empirical relation of \citet{herbst2021} (Eq.~\ref{eq:herbst_contrast}), which modifies the widely used relation of \citet{berdyugina2005} and additionally provides an uncertainty on the predicted temperature.

\begin{equation}
\label{eq:herbst_contrast}
    T_{\rm spot}=-3.58\times10^{-5}T_{\rm eff}^2+1.0188(\pm0.068)T_{\rm eff}+239.3(\pm317.8) ,
\end{equation}

For HD 209458, with $T_{\rm eff}=6091$ K, this relation gives $T_{\rm spot}=5117\pm732$ K. We adopt the central value, $T_{\rm spot,mean}=5117$ K, corresponding to a temperature contrast of $\Delta T_{\rm spot,mean}=974$ K relative to the quiet photosphere. To span the uncertainty in the predicted spot temperature, we use the upper and lower bounds as the maximum and minimum temperatures in our grid: $T_{\rm spot,max}=5849$ K ($\Delta T_{\rm spot,max}=242$ K) and $T_{\rm spot,min}=4385$ K ($\Delta T_{\rm spot,min}=1706$ K), respectively. The minimum spot temperature therefore corresponds to the largest temperature contrast and, consequently, the strongest stellar contamination.

For each spot temperature, we explore five filling factors, $F_{\rm spot}=[0.05,0.1,0.15,0.2,0.25]$. The upper bound of $F_{\rm spot}=0.25$ is motivated by some of the larger spot filling factors inferred from spot-crossing events, such as the polar spot on TOI-3884 \citep{libbyroberts2023,mori2025,murray2026}. Larger filling factors have also been inferred for pre-main-sequence (PMS) stars \citep[e.g.][]{gullysantiago2017,kounkel2024}, potentially covering more than 80\% of the visible stellar disk in the case of the T Tauri star LkCa 4 \citep{gullysantiago2017}. Spot occultations provide stronger constraints on individual spot sizes than inversion techniques applied to unocculted spots, for which strong degeneracies between $T_{\rm spot}$ and $F_{\rm spot}$ are commonly encountered \citep[e.g.][]{thompson2024}.

For the 15 equivalent facula-contaminated spectra, we conservatively adopt temperature contrasts of $\Delta T_{\rm fac}=[100,200,300]$ K, based on measurements of solar faculae \citep{topka1997,gondoin2008,solovev2019}, and use the same filling-factor grid as for the spots, $F_{\rm fac}=[0.05,0.1,0.15,0.2,0.25]$. Facular filling factors can in reality exceed 25\% of the visible stellar disk, particularly for Sun-like stars in which faculae dominate over spots with typical spot-to-facula ratios of $\sim$1:10 \citep[e.g.][]{shapiro2016,chatzistergos2022,nemec2022}. Such large facular coverage is expected to be less common in more active stars, which tend to be spot dominated.

\begin{figure}
    \centering
    \includegraphics[width=\columnwidth]{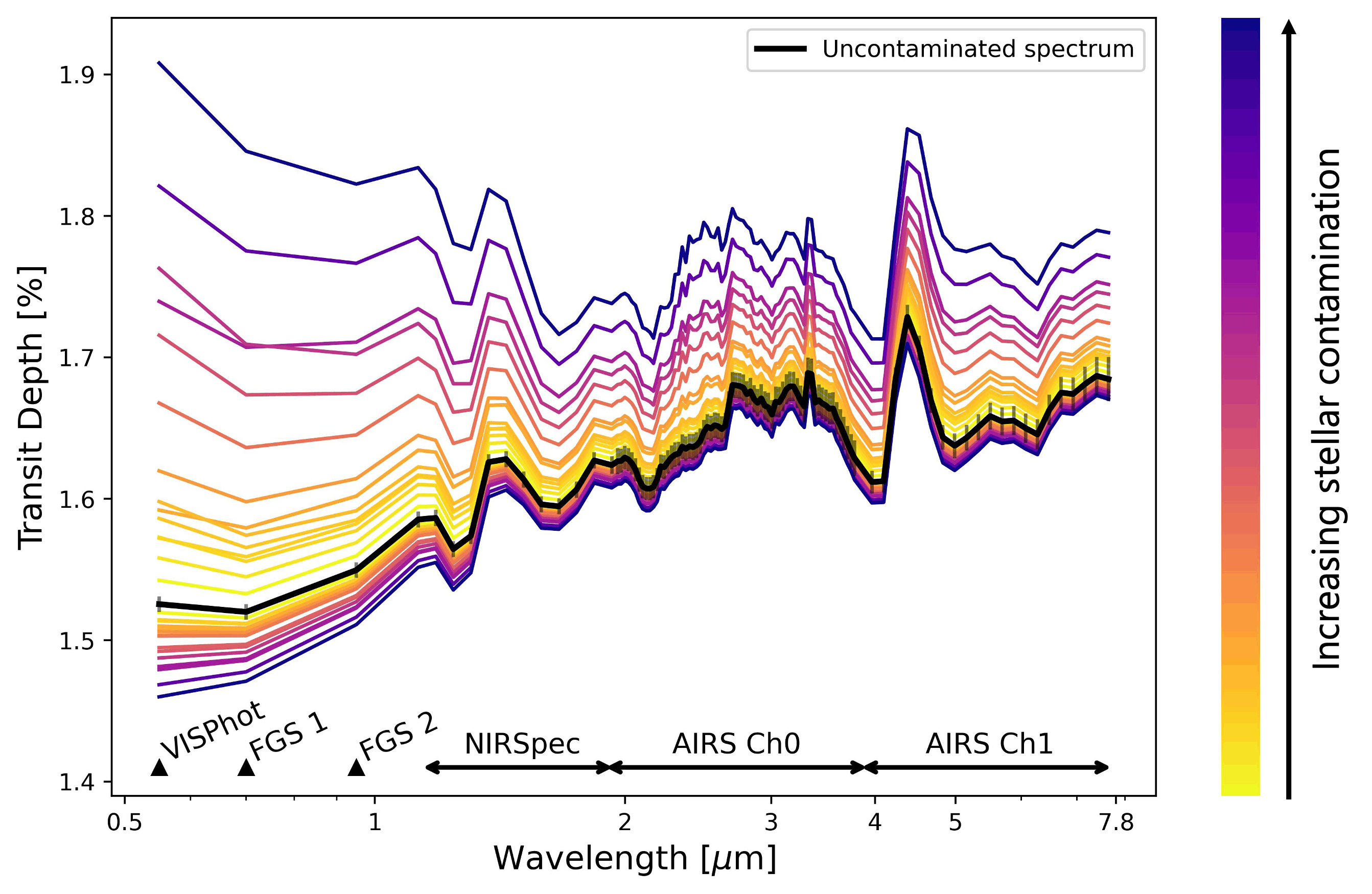}
    \caption{The 30 stellar activity contaminated Ariel spectra generated for this study alongside the equivalent uncontaminated spectrum (\textit{black}). The 15 spectra above the uncontaminated spectrum i.e. characterised by increased transit depths at all wavelengths are spot-contaminated and those below (decreased transit depths) are facula-contaminated. Each forward model possesses the same error bars as the uncontaminated model but these have been removed for clarity. The wavelength ranges of the six Ariel instruments are highlighted along the x-axis.}
    \label{fig:all_cont_spectra}
\end{figure}

The resulting 30 contaminated spectra are shown in Fig.~\ref{fig:all_cont_spectra}. We modelled spots and faculae separately rather than simultaneously because their contamination effects can partially counteract when both are present, reducing the amplitude of the net contamination factor. Isolating the two components therefore allows us to probe the extremes of spot- and facula-driven contamination considered here.

Each contaminated spectrum represents an individual visit to the star–planet system. We then combine subsets of these spectra to simulate the Tier 3 strategy of repeated observations of benchmark systems under variable stellar activity conditions. Three potential approaches to stacking and analysing these visits are explored in the following sections with two main aims: (1) to assess the sensitivity of repeated Ariel observations to stellar variability, and (2) to evaluate different practical approaches for mitigating its impact on the combined observations.

\subsection{Differing Treatments of Multiple Visits}

To understand the consequences of combining visits with varying degrees of inconsistency and how they can be mitigated, we explore three approaches that incorporate progressively greater levels of physical information about stellar variability. 

\textit{Approach 1} is the least physically informed and corresponds to the strategy that would be adopted in the absence of epoch-to-epoch stellar (or planetary) variability i.e. assuming that all visits are mutually consistent within Gaussian noise. In this scenario, the $N$ observed spectra would be stacked without modification and the associated uncertainties assumed to be reduced by a factor of $1/\sqrt{N}$. Approach 1 represents a scalable and readily automated strategy that may be well suited to a homogeneous ‘quick-look’ analysis across the entire Ariel population, but is unlikely to be optimal for the rigorous analysis of individual systems, particularly as stellar activity increases. 

\textit{Approach 2} takes the same unmodified stacked spectra, but instead leverages the highly chromatic nature of stellar contamination to modify the retrieval strategy. Contamination increases rapidly towards shorter wavelengths in the optical (see Fig.~\ref{fig:all_cont_spectra}), whereas infrared contamination has a much weaker wavelength dependence and is more closely approximated by a constant offset: positive for spots and negative for faculae. Approach 2 therefore analyses the infrared, where contamination is generally minimised, separately from the optical by applying data cuts at progressively longer wavelengths and evaluating the resulting response of the retrieval. 

Finally, \textit{Approach 3} explicitly models the stellar contamination effects in each individual visit and corrects for these before combining the spectra. Each Tier 3 visit is expected to achieve a sufficiently high SNR to support a combined star-planet retrieval, which is used to identify the most likely activity parameters, namely the temperature contrasts and filling factors of spots and faculae. These parameters are then used to reconstruct the corresponding wavelength-dependent contamination spectrum, which is subtracted from the individual visit. The corrected visits are subsequently stacked and analysed with a final set of retrievals.

\subsection{Retrieval Set-up}

Retrieval fitting parameters and their associated prior bounds are kept constant across all three approaches and are given in Table~\ref{tab:fitparams}. For each stacked spectrum and each approach we perform paired retrievals: an ASteRA retrieval that accounts for stellar activity and an equivalent baseline retrieval that neglects it. The pairwise set-up facilitates direct model comparison when analysing the results. Combined star-planet retrievals fit for all 9 parameters whereas baseline retrievals do not fit for stellar contamination and therefore have only 5 free parameters. All retrievals use the sampling algorithm MultiNest \citep{feroz2009,buchner2014}, with 1000 live points and a default evidence tolerance of 0.5. There is a small and intentional difference between the forward models and the ASteRA retrievals. Although each individual visit contains contamination from either spots or faculae, both components are always included as free parameters in the retrieval. This is done for two reasons. Firstly, some of the stacked datasets combine spot- and facula-contaminated visits, so using the same set of fitting parameters across all datasets ensures consistency. Secondly, fitting for both components more closely reflects the approach expected for real Ariel observations, where limited or no a priori information on the stellar surface will generally be available, and a broad range of possible activity configurations must therefore be explored if contamination is suspected.

\begin{table}
\centering
\caption{Retrieval fitting parameters and their prior bounds. Activity parameters below the horizontal line are only included in combined star-planet retrievals with ASteRA.} \label{tab:fitparams} 
\begin{tabular}{lcc}
\hline \hline
Fitted Parameter & Prior Bounds   & Scale  \\ \hline
$R_{P}$ [$R_{Jup}$]   & 0.5 ; 3        & linear \\
$T_{P}$ [K]           & 500 ; 3500     & linear \\
log$(H_{2}O)$        & -12 ; -1       & log$_{10}$ \\
log$(CO_{2})$        & -12 ; -1       & log$_{10}$ \\
log$(CH_{4})$        & -12 ; -1       & log$_{10}$ \\ \hline
$T_{spot}$ [K]        & 2500 ; 6080    & linear \\
$F_{spot}$            & 0 ; 0.99       & linear \\
$T_{fac}$ [K]         & 6100 ; 7000    & linear \\
$F_{fac}$             & 0 ; 0.99       & linear \\ \hline
\end{tabular}
\end{table}

\section{Results}
\label{sec:results}

We simulate the combination of four independent visits, approximately corresponding to observations spaced one year apart over Ariel's nominal 4-year mission lifetime. Choosing $N=4$ also provides a convenient benchmark, reducing the formal uncertainties of the stacked spectrum by a factor of two under the incorrect assumption of independent Gaussian noise. We initially generated an ensemble of randomly selected four-visit combinations to explore the broader range of possible visit combinations and their associated contamination levels. From this initial exploration, we selected three representative extreme cases for more extensive retrieval analysis: stacked spectra combining (1) the four most spot-contaminated visits, (2) the four most facula-contaminated visits, and (3) a mixed case combining the two most spot-contaminated and two most facula-contaminated visits.

\subsection{Approach 1: Stacking Multiple Uncorrected Observations}

\begin{figure}
    \centering
    \includegraphics[width=\columnwidth]{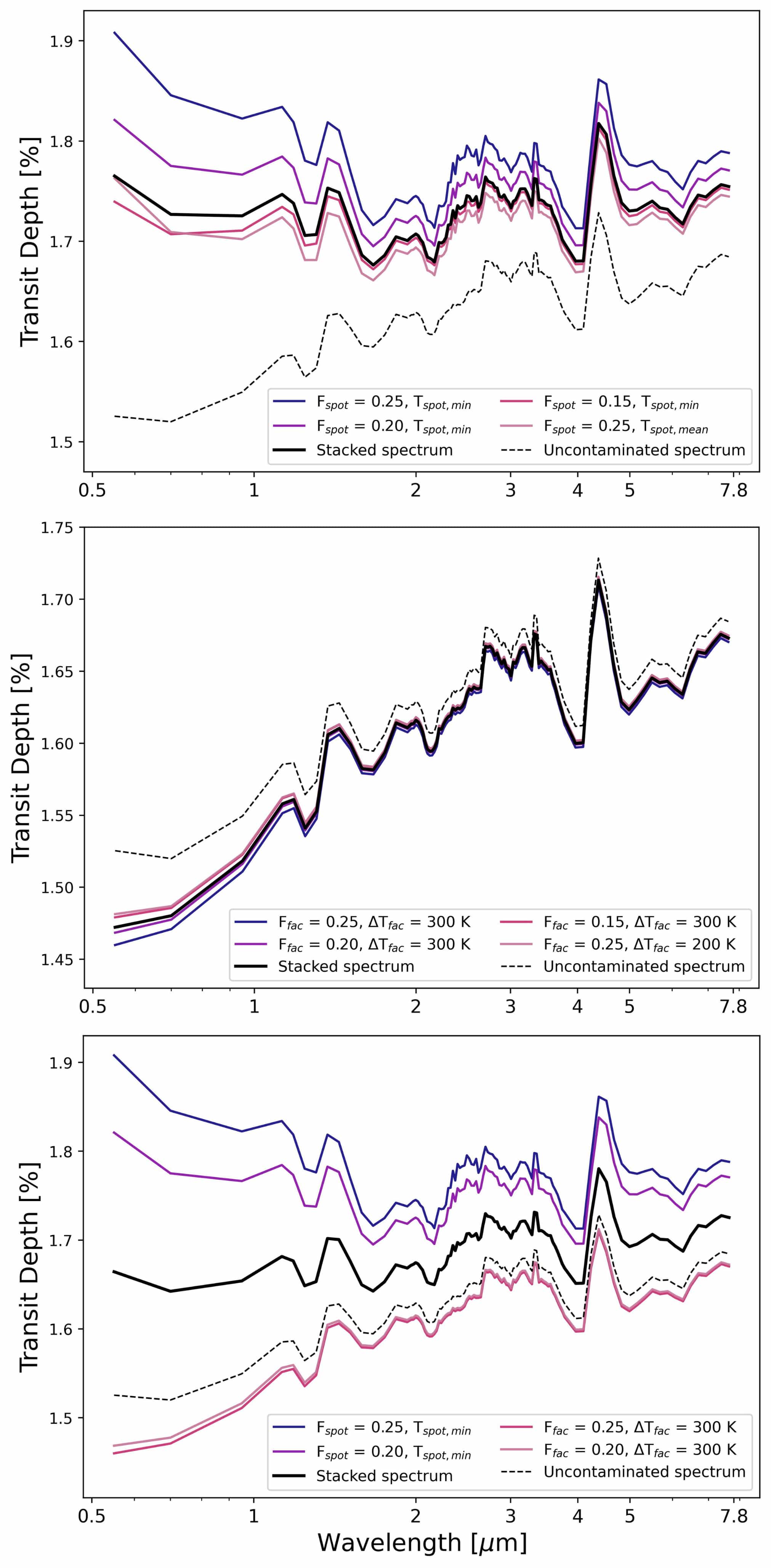}
    \caption{The individual visits and the stacked spectra (\textit{black}) they produce following Approach 1. \textit{Upper Panel} - The combination of the four most spot-contaminated visits. \textit{Middle Panel} - The combination of the four most facula-contaminated visits. \textit{Lower Panel} - The combination of the two most spot-contaminated and two most facula-contaminated visits. The uncontaminated spectrum (\textit{dashed}) is shown in each panel for comparison.}
    \label{fig:spot_fac_stacked_uncorr}
\end{figure}

We expect the primary limitation of this simplisitic approach to be the strong potential for mismatch between its underlying assumptions and the physical reality of a variable host star. As shown in Fig.~\ref{fig:spot_fac_stacked_uncorr}, the resulting stacked spectra are no longer physically representative of any single stellar surface configuration. Instead, they represent the average contribution from four ‘snapshots’ of a time-varying stellar surface. The reduction in error bars can further exacerbate this problem. By implying increased measurement precision, the smaller uncertainties constrain the retrieval more strongly, encouraging it to fit an intrinsically unphysical spectrum with greater statistical confidence. Stress-testing this benchmark approach under conditions of extreme stellar variability therefore provides a useful reference frame to both quantify Ariel's sensitivity to stellar variability and assess the performance of the more rigorous approaches that we subsequently explore.

Retrieved posterior distributions, best-fit spectra, and their residuals are provided in the Appendix in Figures~\ref{fig:spot_uncorr_posterior}, \ref{fig:fac_uncorr_posterior} and \ref{fig:spot_fac_uncorr_posterior} for the three stacked spectra explored, respectively. Tables of the retrieved values and Bayesian evidences are also provided (Tables \ref{tab:spot_results}, \ref{tab:fac_results} and \ref{tab:spot_fac_results}).

The baseline retrievals generally perform poorly in terms of fit to the observed spectrum, the accuracy of the retrieved planetary parameters, and the resulting Bayesian evidence. The stacked spectrum comprising the four most facula-contaminated visits is consistently the least affected, reflecting the substantially smaller amplitude of facular contamination compared with that produced by spots.

The retrieved planetary radius and temperature are particularly strongly affected. Where spot-contaminated visits are included, the radius is overestimated by up to almost $8\%$ in the most strongly affected case, representing a substantial error for a hot Jupiter. Consistent with the observed anti-correlation between these parameters, the temperature is underestimated up to $>50\%$. All molecular abundances are underestimated, with CH$_4$ showing the largest discrepancy, at more than two orders of magnitude. For the stacked spectra containing only facula-contaminated visits, the direction of these biases is reversed: the planetary radius is underestimated, while the remaining planetary parameters are slightly overestimated.

Counterintuitively, the ASteRA retrievals consistently perform better than might be expected given that there is no single set of activity parameters that can perfectly explain and reproduce the observed contamination. Planetary parameters are typically recovered to within 1$\sigma$ of the input values and always within 2$\sigma$. The accuracy of the stellar activity parameters cannot be assessed for the aforementioned reason.

\subsection{Approach 2: Conducting Separate Analyses of the Optical and Infrared Regimes Using Data Cuts}

Approach 2 represents a more physically informed alternative to Approach 1 by exploiting the wavelength dependence of stellar contamination. The infrared is not contamination-free, and retrievals were expected to remain susceptible to biases introduced by the approximately constant offset, particularly in the inferred planetary radius and temperature. However, because the infrared contamination generally exhibits weaker wavelength dependence, it is expected to preserve the overall spectral morphology more closely and introduce less severe biases than its optical counterpart. In the optical, strongly wavelength-dependent contamination can distort genuine planetary absorption features or introduce spurious ones, such as unocculted starspots replicating an apparent Rayleigh-scattering slope. In the case of repeated observations under conditions of stellar variability it is also the regime most impacted by the physical inconsistency introduced when spectra from different stellar configurations are combined.
Analysing the less contaminated infrared region separately may therefore provide a useful compromise between retaining planetary spectral information and reducing sensitivity to stellar contamination.

\begin{figure}
    \centering
    \includegraphics[width=\columnwidth]{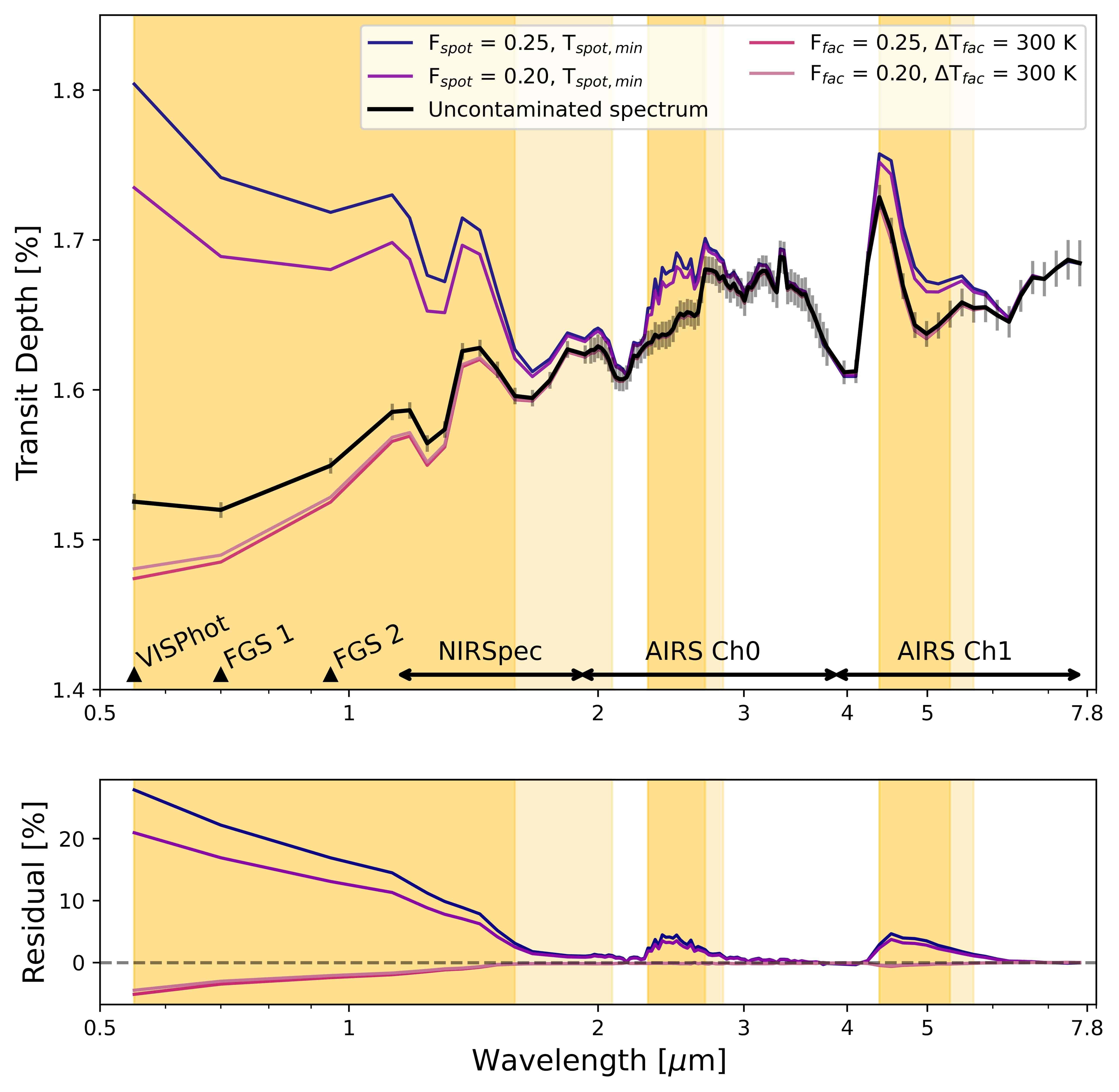}
    \caption{\textit{Upper}: The same contaminated spectra shown in Fig.~\ref{fig:spot_fac_stacked_uncorr} but this time compared to the uncontaminated spectrum (\textit{black}). Offsets have been applied to each contaminated spectrum such that the spectra representing individual visits are aligned with the stacked spectrum at 7.8$\mu$m. \textit{Lower}: The residuals between the offset contaminated spectra and the uncontaminated spectrum. Shaded regions highlight where the spectrum displaying the strongest (spot) contamination (\textit{dark blue}) diverges from the nominal spectrum by upwards of 1\% and 2\% respectively.}
    \label{fig:cont_wav_dependence}
\end{figure}

Figure~\ref{fig:cont_wav_dependence}, however, demonstrates that IR contamination is not entirely wavelength independent. Large, high-contrast spots can introduce broad, shallow features around $\sim$2.5 and 4.5~$\mu$m, which may be associated with CO overtone absorption in the stellar spectrum. These features could subtly enhance genuine planetary absorption in the same wavelength regions or, more concerningly, produce spurious detections, particularly of CO and CO-bearing species such as CO$_2$. Stellar CO contamination has also been observed in high-resolution spectroscopy \citep[e.g.][]{chiavassa2019,genest2022}, demonstrating that stellar molecular features can still contribute to exoplanet spectra across the infrared. Under the strongest activity conditions considered here, Fig.~\ref{fig:cont_wav_dependence} indicates that all Ariel instruments may be susceptible to more complex contamination than a simple offset, although the photometers are disproportionately affected. Conversely, after subtracting the offset component, large regions of the infrared remain within 1\% of the nominal transmission spectrum. This contrast highlights both the potential severity of stellar contamination under unfavourable conditions and the generally greater robustness of the infrared to feature distortion by stellar activity.

\begin{figure}
    \centering
    \includegraphics[width=\columnwidth]{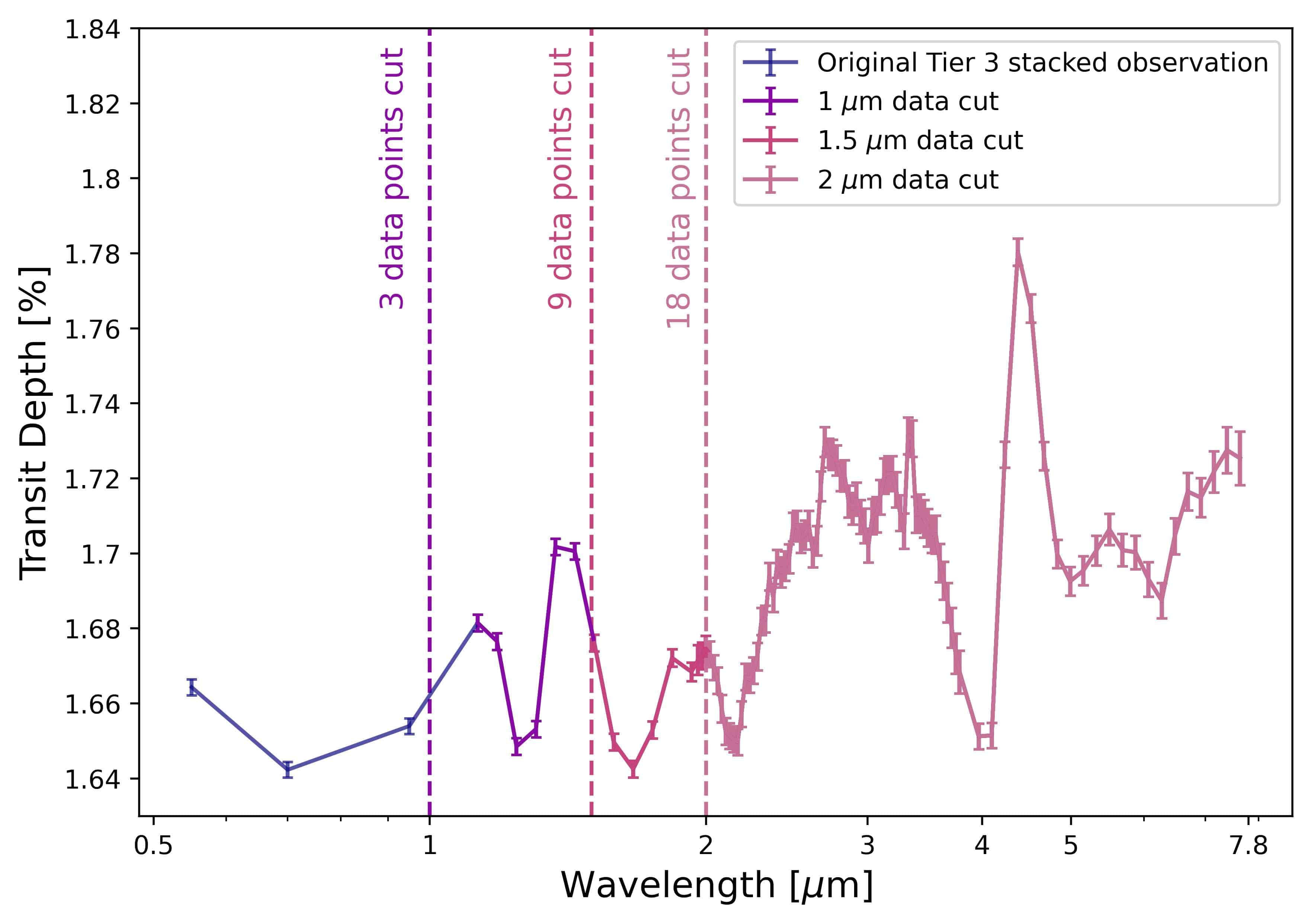}
    \caption{The stacked spectrum that is produced by combining the two most spot contaminated and two most faculae contaminated visits (\textit{blue}) and the truncated spectra resulting from the three datacuts employed at 1 (\textit{purple}), 1.5 (\textit{magenta}) and 2 $\mu$m (\textit{pink}) respectively. The total number of data points removed by each cut is also indicated.}
    \label{fig:optical_IR_datacuts}
\end{figure}

Progressive data cuts are applied to the original stacked Ariel spectrum at 1, 1.5, and 2~$\mu$m. Because of the non-uniform resolution of the Ariel instruments, each successive cut removes an increasing number of data points (Fig.~\ref{fig:optical_IR_datacuts}), resulting in potentially greater information loss as a trade-off for reduced sensitivity to stellar activity. In this sense, Approach 2 follows a similar philosophy to the leave-one-out cross-validation approach of \citet{welbanks2023}, but at a coarser level: rather than assessing the influence of individual data points, we evaluate the information content of broader spectral regions. The purpose of these cuts is to construct a retrieval experiment through which the response of the inferred planetary parameters to the removal of optical information can be directly assessed. Under the assumption that the majority of planetary information is contained in the IR, excluding the more strongly contaminated optical regime should reduce retrieval bias and improve the accuracy of the retrieved planetary parameters up to a point, while incurring only a modest loss in precision.

As a reference case for assessing the information loss associated with each data cut, we first applied Approach 2 to the uncontaminated spectrum. The associated error bars were reduced by a factor of two to match those of the stacked spectra. For this approach, initially only baseline retrievals were conducted, even for the contaminated stacked spectra. Equivalent ASteRA retrievals were conducted later solely to explore the influence of the data cuts on the Bayesian evidence and Bayes factors. If the underlying assumptions of Approach 2 are correct, then we expect to see a decrease in the strength of preference for the ASteRA model as a function of successive data cuts. This behaviour is exactly what was observed and is discussed further in Section~\ref{sec:dis_app2}. Focusing on the impact data cuts have on the baseline retrievals allowed us to directly assess the potential mitigation effect of progressively removing optical information without applying any in-retrieval correction for stellar contamination. The resulting posterior distributions and retrieved spectra for the uncontaminated spectrum are shown in Fig.~\ref{fig:uncont_datacut_posterior}. In this case, successive data cuts have little effect on the accuracy of $R_{\rm p}$ and $T_{\rm p}$, while the associated uncertainties increase only slightly with each cut. The retrieved molecular abundances show somewhat greater sensitivity to the removal of optical information, with the median values becoming increasingly underestimated. However, as the uncertainties also increase, all retrieved values remain well within $1\sigma$ of the input values. Interestingly, the largest loss of precision occurs for the first data cut at $1~\mu$m, despite this removing the fewest data points: only the three photometric bands. The implications of this result are discussed further in Section~\ref{sec:discussion}.

\begin{figure*}
    \centering
    \includegraphics[width=1.8\columnwidth]{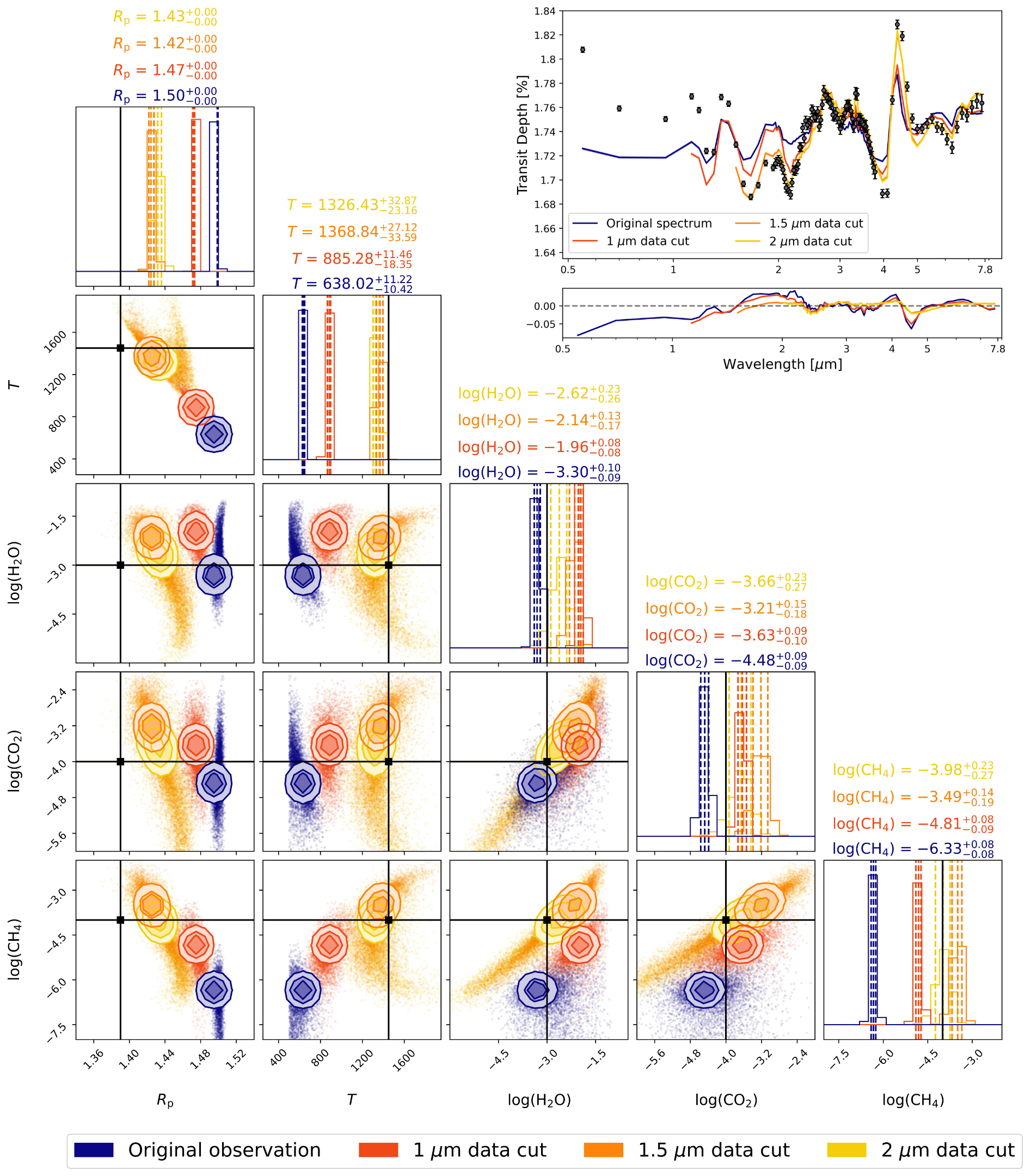}
    \caption{The posterior distributions retrieved when Approach 2 is applied to the stacked spectrum that is consists of spot-contaminated visits only. Retrievals correspond to the original Ariel wavelength range (\textit{blue}), and when data cuts are applied at 1$\mu$m (\textit{red}), 1.5$\mu$m (\textit{orange}) and 2$\mu$m (\textit{yellow}) respectively. Black cross-hairs denote the input, ground truth values. \textit{Inset} - The retrieved best-fit spectra and their residuals.}
    \label{fig:spot_base_datacut_posterior}
\end{figure*}

We then applied the same methodology to the stacked spectra. The four most spot-contaminated visits show the most substantial changes with each data cut, as illustrated in Fig.~\ref{fig:spot_base_datacut_posterior}. The corresponding posterior distributions for the other two stacked spectra are provided in the Appendix (Figures~\ref{fig:fac_datacut_posterior} and \ref{fig:spot_fac_datacut_posterior}). The baseline retrievals over the original wavelength range are identical to those conducted for Approach 1, which were already seen to provide a poor fit to the stacked spectrum, particularly when spot-contaminated visits are included. As successive data cuts are applied, the visual fit to the stacked spectrum improves and the posterior distributions move towards the input ground truth. The accuracy of both $R_{\rm p}$ and $T_{\rm p}$ improves substantially, although some residual inaccuracies remain, which are also explored in Section~\ref{sec:discussion}.

Progressive cuts almost always bring the retrieved $R_{\rm p}$ and $T_{\rm p}$ closer to their input values. The molecular abundances show a more complex response. Initially, as per Approach 1 all abundances are underestimated, although to different degrees, with CH$_4$ showing the largest discrepancy. Applying successive cuts causes the retrieved H$_2$O and CO$_2$ abundances to transition to overestimation at the $1~\mu$m cut, while the same transition occurs for CH$_4$ at $1.5~\mu$m. With the $2~\mu$m cut, all three abundances move back towards their input values. For this cut, the median errors in H$_2$O and CO$_2$ are comparable in magnitude to those obtained for the original spectrum, but their direction is reversed. However, the associated uncertainties also increase, bringing both retrieved values within $2\sigma$ of the input values. The largest improvement is seen for CH$_4$, for which the input abundance is recovered almost exactly. This improvement in the CH$_4$ abundance means that the most accurate interpretation of the overall atmospheric chemistry is obtained when the 2$\mu$m data cut is applied.

The results presented here show that excluding the more strongly contaminated optical region incurs only a modest loss in retrieval precision while substantially improving the accuracy of the retrieved planetary parameters, most notably the chemical abundances. 

 \subsection{Approach 3: Correcting Individual Visits Prior to Stacking}

Because each Tier 3 visit is designed to achieve a sufficiently high SNR individually, a more physically accurate approach is to correct each visit for stellar contamination prior to stacking. For each visit, we perform a combined star–planet retrieval and use the retrieved stellar activity parameters to reconstruct the best-fitting contamination factor as a function of wavelength, which is then subtracted from the observed spectrum. Although the stellar activity parameters can exhibit substantial degeneracies, this should not necessarily compromise the correction. Indeed, the degeneracy arises because different combinations of spot and facula properties can produce very similar wavelength-dependent contamination factors. The quality of the correction therefore relies on accurately recovering the contamination spectrum, rather than uniquely constraining the underlying stellar surface configuration.

\begin{figure}
    \centering
    \includegraphics[width=\columnwidth]{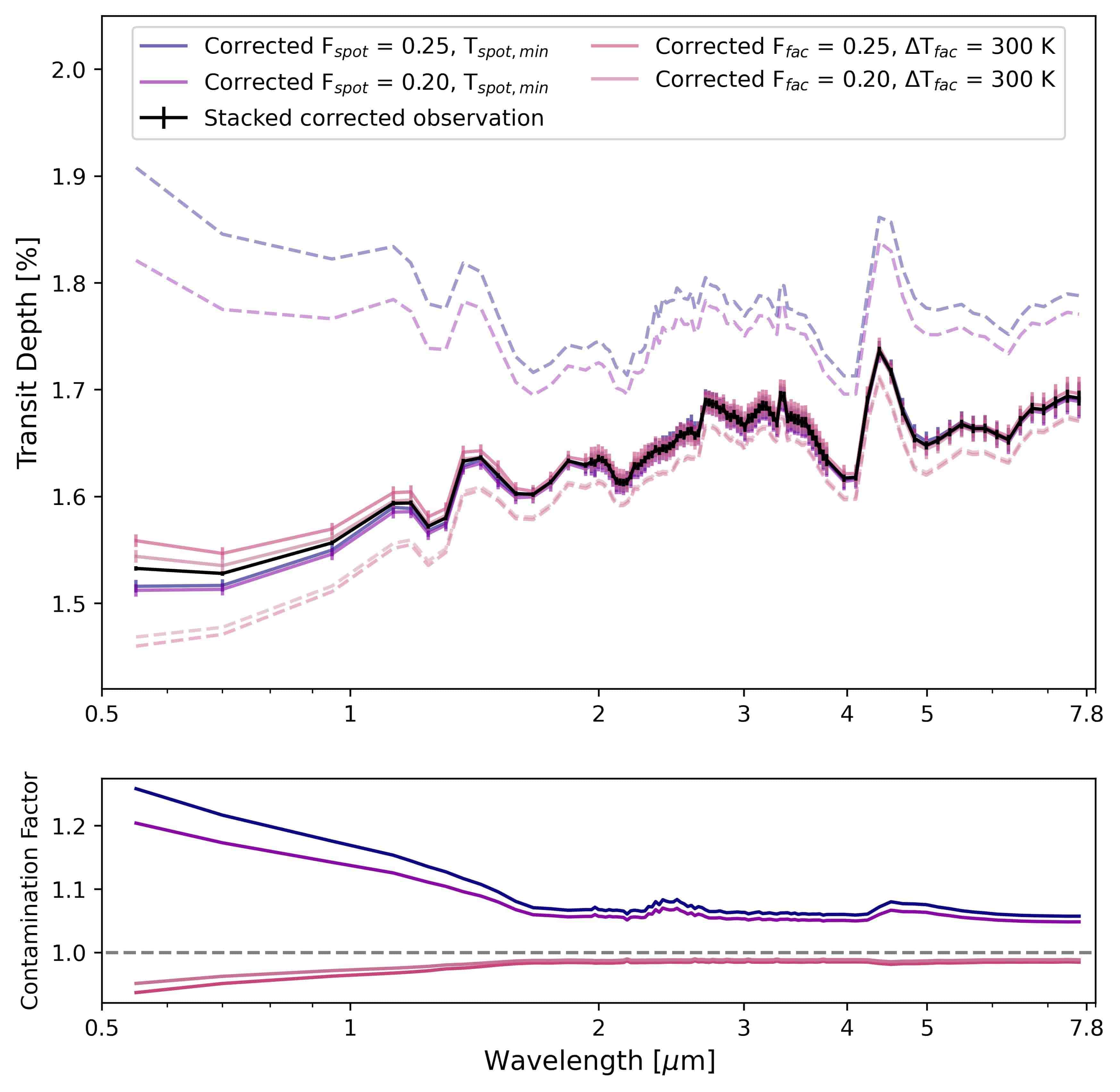}
    \caption{\textit{Upper}: The two most spot- and facula-contaminated visits pre- (\textit{dashed}) and post-activity correction (\textit{solid}) with ASteRA, the black line denotes the stacked spectrum produced by combining the four corrected spectra. \textit{Lower}: The contamination factors as a function of wavelength retrieved by ASteRA that are subtracted to correct each spectrum.}
    \label{fig:spot_fac_corr_stacked}
\end{figure}

Approach 3 is considered to be successful if it returns corrected spectra that are substantially more consistent with one another than they were initially. Assuming that there is negligible planetary variability, this is an indication that ASteRA has accurately reproduced and corrected the associated contamination spectra. Improved visit-to-visit consistency allows us to overcome the primary limitation of Approach 1 without sacrificing information as required for Approach 2. This is demonstrated by the four corrected individual visits shown in Fig.~\ref{fig:spot_fac_corr_stacked}. The corrected spectra are substantially more consistent with one another than the original contaminated visits, with near-perfect agreement in the infrared. Minor discrepancies remain in the optical, however, corresponding to a slight overcorrection of the contamination in each case which is examined in detail in Section~\ref{sec:dis_app3}. This improved consistency, however, comes at a significant computational cost. The method requires $N+1$ retrievals: one ASteRA retrieval for each of the $N$ visits to model and remove the stellar contamination, followed by a final retrieval of the stacked, corrected spectrum. Providing that the individual corrections are sufficiently accurate, the final retrieval should no longer require an explicit stellar activity model. The computational cost therefore scales directly with the number of visits and is higher than for either alternative. Approach 1 requires only a single retrieval, while Approach 2 requires as few as two retrievals depending on the number of wavelength cuts considered, although for completeness four retrievals were conducted to explore three potential data cuts. Approach 3 therefore provides the most physically consistent treatment of stellar variability, but at the expense of substantially greater computational cost.

Retrieved posterior distributions comparing the performance of Approach 3 against Approach 1 can be found in the Appendix (Figures~\ref{fig:spot_approach3}, \ref{fig:fac_approach3} and \ref{fig:spot_fac_approach3}).

The uncertainties on the retrieved planetary parameters with Approach 3 are consistently smaller than those obtained with Approach 1 when a single activity correction is applied. For the stacked spectrum containing both spot- and facula-contaminated visits, Approach 3 performs particularly well, recovering all planetary parameters to within $1\sigma$, with highly precise associated uncertainties (Fig.~\ref{fig:spot_fac_approach3}). In contrast, for the stacked spectra containing exclusively spot- or facula-contaminated visits, the effects of the slight overcorrections accumulate, resulting in retrieved parameters that are generally slightly less accurate than those obtained with the single correction applied in Approach 1. The improved performance for the mixed spot-facula case therefore suggests that the marginal overcorrections associated with the two types of stellar contamination partially canceled out one another when combined. The likely origin of this contrasting behaviour is also considered in more detail in Section~\ref{sec:dis_app3}.

\section{Discussion}
\label{sec:discussion}

\subsection{Performance of the Three Approaches}
\subsubsection{Approach 1}

For a suite of approaches such as those outlined here to be useful for Ariel, it is important to understand both where they perform well and the physical reasons underlying their performance. Approach 1 performs better than might be anticipated given that it does not explicitly account for stellar variability. Its relatively strong performance can likely be attributed, at least in part, to the non-uniform spectral resolution of Ariel's instruments. The unphysical nature of the stacked spectrum is most pronounced in the optical regime $\leq 1.7~\mu$m, where the contamination is more closely represented by a power-law. The strongest contamination effects are therefore concentrated primarily within the three photometric bands, with some contribution to the NIRSpec spectrometer coverage. At longer wavelengths, the offset-like contamination is comparatively easier for the retrieval to accommodate, despite the underlying inconsistency of the stacked stellar spectrum. These wavelengths also contain the bulk of the planetary information and are covered by Ariel's highest-resolution spectrometers, particularly AIRS Ch0. The greater spectral resolution provides more information with which to constrain the planetary atmosphere and weights the retrieval more heavily towards achieving the best fit to this region, facilitating the accurate recovery of the molecular abundances. 

Additionally, the strong performance for the stacked spectra containing exclusively spot- or facula-contaminated visits may be linked to the activity parameters characterising the individual visits. For the stacked spectrum containing only spot-contaminated visits, three of the four visits have the same spot temperature, $T_{\rm spot,min}=4385$ K, while the remaining visit has $T_{\rm spot,mean}=5117$ K. This may reduce the ambiguity in the retrieval of the spot temperature, with the retrieved value of 4848 K lying almost exactly between the two input temperatures. The larger amplitude of the averaged contamination is instead accommodated by an increased spot filling factor, with the retrieval returning $F_{\rm spot}=0.40^{+0.10}_{-0.09}$, exceeding the maximum filling factor imposed for any individual visit ($F_{\rm spot}=0.25$). This provides a further illustration of the degeneracy between stellar activity parameters, which can become particularly difficult to break in low resolution spectroscopy. A similar behaviour is seen for the stacked spectrum containing only facula-contaminated visits, where again three of the four individual visits share the same temperature contrast ($\Delta T_{\rm fac}=300$ K). In this case, both the retrieved facular temperature contrast ($\Delta T_{\rm fac}=533$ K) and filling factor ($F_{\rm fac}=0.61^{+0.19}_{-0.27}$) are overestimated to an even greater extent. This may be partly related to the fact that all retrievals allow for contributions from both spots and faculae. The resulting parameter degeneracies permit solutions in which spot contribution is present but is largely counteracted by a simultaneous increase in the facular contribution, producing a similar net contamination spectrum.

\subsubsection{Approach 2}
\label{sec:dis_app2}

\begin{figure*}
    \centering
    \includegraphics[width=2\columnwidth]{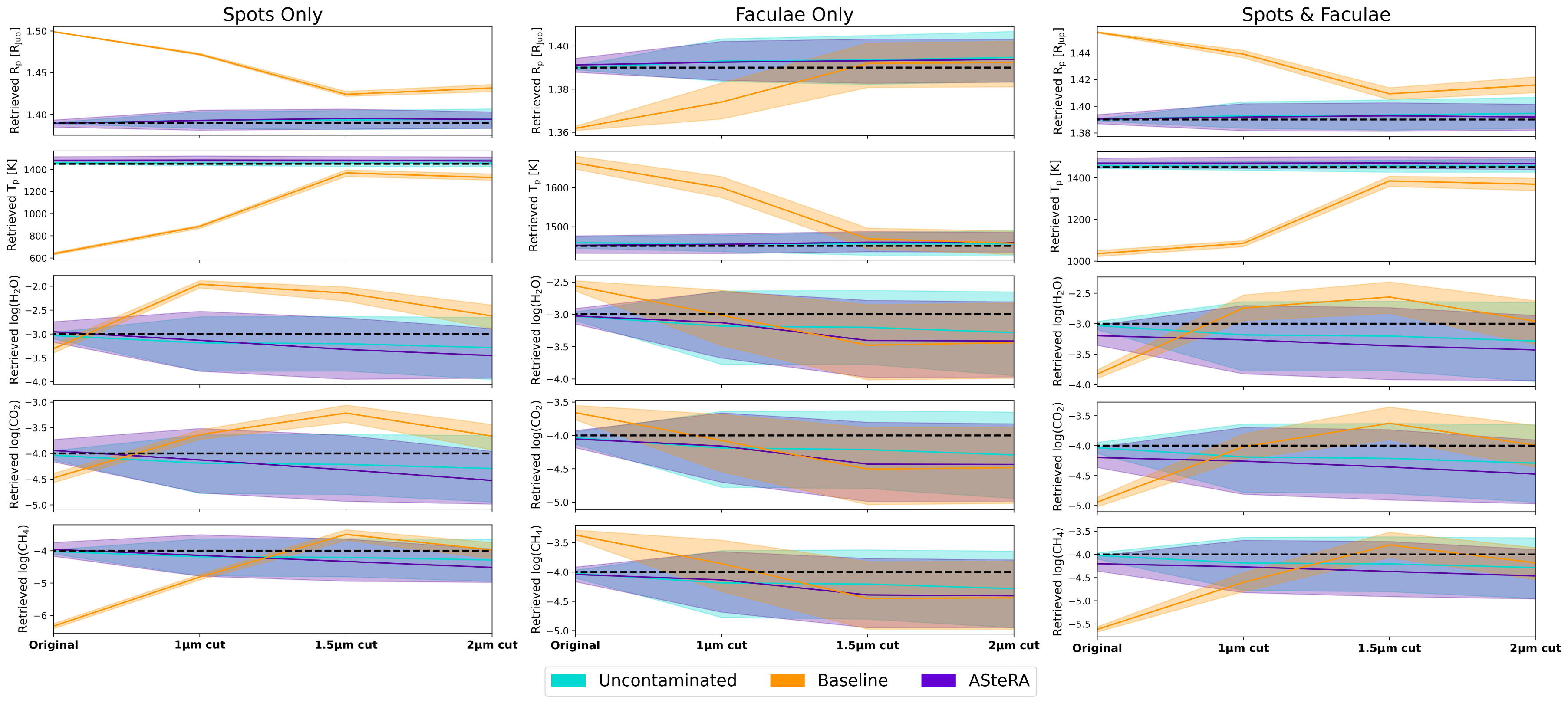}
    \caption{Summary plots showing the retrieved planetary parameters and their uncertainties as a function of applying consecutive data cuts to the spot case only stacked spectrum (\textit{left}), the facula case only stacked spectrum (\textit{middle}) and the case where spot- and facula-contaminated visits are combined (\textit{right}). The retrieval experiment is conducted on the stacked spectra using a baseline model (\textit{orange}) and with the ASteRA model (\textit{purple}) for comparison. The same methodology is also applied to an uncontaminated spectrum (\textit{teal}) to act as a reference frame for how the information loss affects the precision of the retrieved parameters.}
    \label{fig:summarystats}
\end{figure*}

Summary plots for all of the retrievals conducted under Approach 2 are shown in Fig.~\ref{fig:summarystats}. Overall, the retrievals show that applying successive data cuts substantially improves the accuracy of the retrieved parameters in exchange for a modest loss of precision, and without requiring an in-retrieval correction for stellar activity. This is because the information that was removed was most strongly affected by the missing stellar physics that the baseline retrieval model could not account for. To facilitate Bayesian model comparison, we conducted additional retrievals following the same approach but this time with ASteRA to account for stellar contamination that remains in the cropped spectra. As expected, the ASteRA models are consistently preferred, but the Bayesian evidence also shows that the preference for the more complex model decreases as the most strongly contaminated optical wavelengths are removed. Across the nominal Tier 3 wavelength range, the ASteRA model is preferred over the worst-performing baseline model by a Bayes factor of 2863.8, whereas after the final data cut this preference decreases by an order of magnitude to 224.3 (Table ~\ref{tab:spot_results}). Whilst both values represent a conclusive preference for the ASteRA model, the substantial reduction in the Bayes factor indicates that the preference is driven disproportionately by the optical region. For the stacked spectrum with facula-only contamination, the Bayesian evidence for the two models is essentially indistinguishable by the second data cut, with $\ln(\mathscr{B})=0.3$, indicating no significant preference for either model (Table ~\ref{tab:fac_results}).

The comparative effect of each data cut can be most readily interpreted in terms of the morphology of the contamination spectrum. For $R_{\rm p}$ and $T_{\rm p}$, cuts at 1 and 1.5~$\mu$m produce substantial improvements in accuracy, whereas a further cut at 2~$\mu$m provides little to no additional benefit. This is likely because the contamination transitions to approximately offset-like within the 1.5 - 2$\mu$m range , such that removing additional spectral information no longer has an accompanying mitigation effect. Other physical processes, such as clouds, can also produce similar infrared offsets and may therefore be degenerate with stellar contamination, potentially complicating more realistic retrieval experiments. The offset also explains why, in the absence of an in-retrieval correction, Approach 2 improves the accuracy of $R_{\rm p}$ and $T_{\rm p}$ but cannot completely remove their biases, whereas the equivalent retrieval with ASteRA can (Fig.~\ref{fig:summarystats}). When spot contamination is present, the radius remains slightly overestimated and the temperature slightly underestimated, as the retrieval accommodates the residual offset through adjustments to these parameters. When only facular contamination is present, data cuts at 1.5 and 2~$\mu$m recover $R_{\rm p}$ and $T_{\rm p}$ with an accuracy comparable to that obtained for the uncontaminated spectrum and for the corresponding retrievals with an in-retrieval activity correction. This is likely due to the relatively low amplitude of facular contamination in the infrared, such that the resulting offset has little effect on the retrieved planetary parameters.

The behaviour of the molecular abundances is more difficult to interpret, particularly given their transition from under- to overestimation after removing only the three photometric bands. The corresponding experiment on the uncontaminated spectrum provides useful context: the largest relative loss of precision occurs following the 1~$\mu$m data cut, suggesting that these three broad photometric bands provide important constraints on the overall spectral shape. Their inclusion may help establish the underlying continuum, including features such as the presence and extent of a Rayleigh-scattering slope and the depth at which the atmosphere becomes opaque. The subsequent overestimation of the molecular abundances may be related to the residual spot contamination, which increases the observed transit depth across the affected wavelengths. The baseline retrieval may then compensate for these enhanced transit depths by increasing the retrieved chemical volume mixing ratios as well as by fine-tuning $R_{\rm p}$ and $T_{\rm p}$. 

Fig.~\ref{fig:summarystats} also shows that the uncertainties on the baseline model are almost always narrower than those obtained with the other two models. When compared to the ASteRA retrieval, this is known a consequence of model oversimplification and incompleteness as the posterior uncertainties are conditional on the parameter space defined by the retrieval itself \citep[e.g.][]{rocchetto2016,nixon2024,barstow2026}. The baseline model has both a low dimensionality, with only five free parameters, and an incomplete description of the underlying physical processes, as the additional degrees of freedom required to account for stellar contamination are not accessible to it. Within this restricted parameter space, only a relatively narrow range of parameter values may provide the best fit to the observations, resulting in apparently precise constraints. However, these constraints do not account for the uncertainty associated with the missing stellar physics, and the retrieved parameters can therefore be precise without being accurate.The results of Approach 2 applied to the uncontaminated spectrum demonstrate that this increased precision cannot be attributed to the lower dimensionality of the baseline model alone. When the same five-parameter retrieval model is applied in the absence of stellar contamination, the model can accurately explain the observation, and a comparatively broader range of parameter values can provide a good fit, resulting in wider posterior uncertainties.

Information loss is not generally desirable, particularly for an instrument suite such as Ariel that is specifically designed to provide broad wavelength coverage across both the optical and infrared. Nevertheless, separating the analysis of these two wavelength regimes provides a practical advantage: the infrared can retain much of the planetary information while reducing sensitivity to the more strongly wavelength-dependent component of stellar contamination. Through Approach 2, much of this information can be accessed without requiring a combined star-planet retrieval and its associated higher dimensionality. This may be particularly useful for Ariel, where retrievals will already need to account for a wide range of physical processes within high-dimensional parameter spaces, regardless of whether stellar activity is explicitly modelled or not.
The progressive data-cut framework could also provide a useful diagnostic for assessing potential stellar contamination when its presence is less certain. In the absence of contamination, applying successive optical data cuts produces relatively little change in the retrieved planetary parameters (Fig.~\ref{fig:uncont_datacut_posterior}), whereas contaminated spectra show systematic changes in the retrieved parameters as the optical information is progressively removed, with the results eventually tending towards the input values (Fig.~\ref{fig:spot_base_datacut_posterior}). Although this approach is computationally more demanding than applying a single retrieval to the full wavelength range, its use alongside other approaches and activity mitigation techniques could provide an independent check on the robustness of the retrieved planetary properties and increase confidence in the interpretation of potentially activity-contaminated observations.

\subsubsection{Approach 3}
\label{sec:dis_app3}

As the highest-fidelity model, Approach 3 was expected to provide the most rigorous treatment of stellar variability and, consequently, the largest improvements in both retrieval accuracy and precision. While the approach performs well overall, the results also highlight an imperfect implementation of the individual activity corrections. For each individual visit only spot or facular contamination is present, however, both components are included as free parameters in the retrieval. The absence of either component can be represented either by converging towards the lower bound of its filling factor, $F = 0$, or by converging towards a feature temperature that is effectively identical to that of the quiet photosphere, corresponding to the upper bound for $T_{\rm spot}$ and the lower bound for $T_{\rm fac}$. A minimum filling factor of zero is physically motivated, as negative filling factors are unphysical. However, imposing hard prior bounds can introduce boundary effects \citep{buchner2026}. In the case of the filling factors, the posterior is truncated at zero, meaning that MultiNest can only sample the range of positive spot or facular contributions that remain compatible with the data. Consequently, the posterior for an absent activity component is inherently asymmetric and cannot be represented by a symmetric uncertainty around the true value of zero. This can result in a small positive contribution being inferred for the activity component that is not present, producing the slight over-corrections observed. Further evidence that the performance is hindered by boundary effects is that the best performance is achieved for the spectrum combining both spot- and facula-contaminated visits. In this scenario, the over-corrections in the individual visits partially oppose one another, resulting in a corrected spectrum that is closer to the uncontaminated case.

\subsection{Scope and Future Directions}

The simulations presented here necessarily provide a simplified representation of the observations that Ariel may obtain under conditions of stellar variability, particularly in terms of the physical complexity of the stellar and planetary models. The 1D stellar activity model implemented in ASteRA provides a useful framework for describing the wavelength-dependent effects of spots and faculae, but may not capture the full complexity of stellar contamination in real observations. For the most active stars, higher-fidelity approaches, including 3D stellar models, may ultimately be required to describe and correct more complex surface structures and their associated spectral signatures. Such models allow us to account for spatially dependent effects such as limb darkening and centre-to-limb variations, as well as the influence of localised magnetic fields and associated magnetohyrodynamic (MHD) processes. These considerations are likely to be particularly important for faculae, for which 1D stellar models may provide an inadequate representation \citep{norris2017,norris2023,witzke2022}.
The planetary model is similarly idealised, with a small number of trace species present at relatively high volume mixing ratios and no additional scatter imposed on the simulated binned data points. Other sources of atmospheric opacity, such as clouds, are also not considered at this stage and could introduce additional degeneracies with the stellar activity parameters. These choices were made to isolate the effects of variable stellar contamination in a controlled setting.

The range of stellar variability explored here was also intentionally chosen to represent an extreme scenario, particularly for the transitions between strongly spot- and facula-dominated visits. Such transitions are observationally motivated but over substantially longer timescales: young, active stars tend to be more spot-dominated, with the relative contribution of faculae increasing as stars age and spin down \citep{shapiro2016, reinhold2019, nemec2022}. On shorter timescales, changes in the observed balance of spots and faculae may also arise as active regions rotate into and out of view. However, the magnitude of variability produced by such changes is unlikely to reach the full range explored in these simulations. The adopted configurations should therefore be regarded as a stress test of the analysis approaches rather than a representation of the variability expected for typical Ariel targets.
Within this experiment framework, our simulations provide a proof of concept for exploring how stellar variability may imprint itself on Ariel spectra and for testing complementary approaches under consistent conditions. This enables a direct assessment of how stellar contamination can be identified, characterised, and mitigated across repeated observations. Future work can and should build on these results by introducing more realistic stellar and planetary models and exploring a broader range of activity amplitudes, stellar types, and temporal variability. A further, more physically motivated “Approach 4” would be to develop a retrieval framework in which all visits are analysed simultaneously, while allowing the photospheric configuration to vary between visits. Such a framework is already under development and will be the subject of future work.

A key challenge is how to interpret systems in which both the planet and star are variable, where there is a risk of inadvertently attributing stellar variability to the planet, or vice versa. In this scenario, independent complementary observations will likely be essential for disentangling the two competing signals, including simultaneous multi-band photometry from the ground and precise constraints on stellar parameters likely to influence variability, such as age and rotation period. Long-baseline stellar light curves from TESS and PLATO will also provide constraints on the variability amplitudes of many Ariel target stars, informing decisions about the need for preparatory and contemporaneous monitoring.

\section{Conclusions}
\label{sec:conclusions}
Ariel’s Tier 3 observations will offer a unique opportunity to study how both planets and their host stars vary over time, providing insights that extend beyond the mission and across the wider exoplanet community. This work represents a pilot study that explored how sensitive the Ariel tiered observing strategy could be to stellar variability on a visit-to-visit basis and introduced three practical approaches of varying complexity that can be used to understand the impact of this variability and mitigate for it. Understanding the potential magnitude of stellar variability contributions in spectroscopy is an important first step towards maximising the science return of Tier 3 and studying planets and their host stars as coupled systems evolving within a shared environment. The key findings of this work are summarised below.

\begin{enumerate}
    \item Stellar variability will be an important consideration for the Ariel observational strategy, as multiple visits are a core component of every tier: in Tiers 1 and 2, to build up the required SNR, and in Tier 3, to search for planetary variability within the highest SNR benchmark targets. Repeated observations of the same star-planet system cannot necessarily be treated as statistically independent measurements of a single underlying spectrum if the stellar surface evolves substantially between observations.
    \\
    \item The naive, unmodified stacking methodology of Approach 1 provides a useful reference case but should not be considered sufficient for the robust analysis of highly variable stars. Despite the physical inconsistency introduced by combining different stellar configurations, a single stellar activity correction with ASteRA can still recover the planetary parameters reasonably well. However, it may not perform adequately under all variability conditions and also should not be interpreted as evidence that the impact of stellar variability is negligible.
    \\
    \item Applying progressive optical data cuts following Approach 2 provides a complementary diagnostic that does not require an in-retrieval activity correction. Our retrieval experiments show that, despite the associated loss of spectral information, the data cuts can substantially improve the accuracy of the retrieved planetary parameters up to a point, at the expense of only a modest reduction in precision. The response of the retrieved parameters to the successive data removal can also be an indicator of stellar contamination by itself, which may be particularly useful in situations where the activity of the host star is uncertain.
    \\
    \item Correcting individual visits before combining them provides the most physically realistic treatment of temporal stellar variability, but at substantially greater computational cost. The approach generally produces more consistent corrected spectra and smaller uncertainties, although small overcorrections can arise from boundary effects when an activity component is absent but remains a free parameter in the retrieval.
    \\
    \item Developing a flexible suite of variability mitigation approaches is well suited to Ariel's large and diverse target population, enabling the highest fidelity, computationally intensive methods to be reserved for cases where simpler approaches are insufficient. This can help us push towards maximising both the efficiency and robustness of large-scale atmospheric characterisation.
    \\
    \item For particularly challenging cases, multiple approaches can and should be used in combination as they provide distinct ways of assessing and tackling the impact of stellar variability. Although the approaches presented here cannot be considered truly independent as they share underlying models and assumptions, they provide complementary cross-checks on the robustness of the retrieved planetary properties.

\end{enumerate}

Together, these results demonstrate that stellar activity and variability should be considered when selecting and prioritising Ariel targets, but need not prevent the atmospheric characterisation of their planets when adequately accounted for. Retaining a representative sample of active systems, including within Tiers 2 and 3, would be consistent with Ariel's goal of characterising a diverse range of planets and could also facilitate unique science cases, such as investigating star-planet interactions and the atmospheric response to stellar activity e.g. flares and potentially even activity cycles. Preparatory and contemporaneous monitoring, particularly through long-timescale, multi-band photometry, would be especially valuable for characterising the activity and variability of potential targets and informing the appropriate observing and analysis strategies. This would help ensure that active systems can be included in the final Ariel MRS whilst minimising the associated challenges in their atmospheric characterisation.

\section*{Acknowledgements}

A.T. postdoctoral fellowship was funded by the European Space Agency, through  ESA Contract N. 4000146542/24/ES/CM. A.T. also acknowledges support from the ASI-INAF agreement 2021-5-HH.2-2024.

\section*{Data Availability}

This work has utilised publicly available datasets such as the PHOENIX BT-Settl models which are hosted by the Spanish Virtual Observatory (https://svo.cab.inta-csic.es/) and open-source software including TauREx 3 \citep[][https://github.com/ucl-exoplanets/taurex3]{alrefaie2021} and MultiNest \citep{feroz2009}/PyMultiNest \citep{buchner2014}.



\bibliographystyle{rasti}
\bibliography{bibliography}




\appendix

\section{Appendix}

\begin{table*}
\label{tab:spot_results}
\caption{Retrieval results for the three approaches applied to the spot-contaminated stacked spectrum. ln$\mathscr{B}$ is the Bayesian evidence returned and the Bayes factors comparing baseline and ASteRA model pairs are given next to the preferred model. Note that the Bayesian evidence should only be compared between the ASteRA and baseline retrievals for a given approach and not across the approaches as they have different input spectra.}
\centering
\begin{tabular}{lllllll}
\hline \hline
                   & $R_{\rm p}$ [$R_{\rm Jup}$] & $T_{\rm p}$ [K] & log(H$_2$O) & log(CO$_2$) & log(CH$_4$) & ln$\mathscr{B}$ (\textbf{Bayes Factor}) \\ \hline
\multicolumn{7}{l}{\textbf{Approach 1}}                                                                             \\ \hline
Baseline           & 1.499 $\pm 0.000$  & 638 $^{+11}_{-10}$  & -3.30 $^{+0.10}_{-0.09}$ &  -4.48 $\pm 0.09$ & -6.33 $\pm 0.08$  & -1963.2 $\pm$ 0.2
                \\
ASteRA             & 1.389 $\pm 0.004$  &  1482 $^{+32}_{-23}$  & -2.95 $^{+0.21}_{-0.22}$ & -3.94 $^{+0.21}_{-0.22}$ &  -3.96 $^{+0.22}_{-0.21}$ & 900.6 $\pm$ 0.2 \textbf{(2863.8)}               \\ \hline
\multicolumn{7}{l}{\textbf{Approach 2}}                                                                             \\ \hline
Baseline (1 $\mu$m)  & 1.472 $\pm 0.001$ & 885 $^{+11}_{-18}$  & -1.96 $\pm 0.08$ & -3.63 $^{+0.09}_{-0.10}$ & -4.81 $^{+0.08}_{-0.09}$ & -479.5 $\pm$ 0.2                \\
ASteRA (1 $\mu$m)    & 1.393 $\pm 0.012$ & 1484 $^{+38}_{-28}$ & -3.13 $^{+0.61}_{-0.65}$ & -4.12 $^{+0.61}_{-0.64}$ & -4.14 $\pm 0.64$ & 876.0 $\pm$ 0.2      \textbf{(1355.5)}         \\
Baseline (1.5 $\mu$m) & 1.424 $^{+0.004}_{-0.002}$ & 1369 $^{+27}_{-34}$ & -2.14 $^{+0.13}_{-0.17}$ & -3.21 $^{+0.15}_{-0.18}$ &  -3.49 $^{+0.14}_{-0.19}$ & 537.9 $\pm$ 0.1                \\
ASteRA (1.5 $\mu$m)  & 1.395 $^{+0.011}_{-0.013}$ & 1483 $^{+34}_{-27}$ & -3.32 $^{+0.66}_{-0.62}$ & -4.31 $^{+0.67}_{-0.61}$ &  -4.33 $^{+0.69}_{-0.61}$ & 820.8 $\pm$ 0.2  \textbf{(282.9)}              \\
Baseline (2 $\mu$m)  & 1.432 $\pm 0.004$ & 1326 $^{+33}_{-23}$ & -2.62 $^{+0.23}_{-0.26}$ & -3.66 $^{+0.23}_{-0.27}$ & -3.98 $^{+0.23}_{-0.27}$ & 520.9 $\pm$ 0.1                \\
ASteRA (2 $\mu$m)    & 1.394 $^{+0.008}_{-0.010}$ & 1477 $^{+36}_{-29}$ & -3.45 $^{+0.57}_{-0.47}$ & -4.52 $^{+0.57}_{-0.46}$ & -4.52 $^{+0.57}_{-0.46}$ & 745.2 $\pm$ 0.2 \textbf{(224.3)}               \\ \hline
\multicolumn{7}{l}{\textbf{Approach 3}}                                                                             \\ \hline
Baseline           & 1.380 $\pm 0.001$ & 1540 $\pm 14$ & -2.78 $\pm 0.07$ & -3.76 $^{+0.10}_{-0.09}$ & -3.71 $\pm 0.07$ & 909.9 $\pm$ 0.2                \\
ASteRA             & 1.392 $^{+0.003}_{-0.004}$ & 1469 $^{+26}_{-21}$  & -3.08 $^{+0.13}_{-0.14}$ & -4.05 $^{+0.14}_{-0.15}$ & -4.09 $^{+0.14}_{-0.15}$  & 914.3 $\pm$ 0.2 \textbf{(4.4)}               \\ \hline
\end{tabular}
\end{table*}

\begin{table*}
\label{tab:fac_results}
\caption{Retrieval results for the three approaches applied to the facula-contaminated stacked spectrum. ln$\mathscr{B}$ is the Bayesian evidence returned and the Bayes factors comparing baseline and ASteRA model pairs are given next to the preferred model.}
\centering
\begin{tabular}{lllllll}
\hline \hline
                   & $R_{\rm p}$ [$R_{\rm Jup}$] & $T_{\rm p}$ [K] & log(H$_2$O) & log(CO$_2$) & log(CH$_4$) & ln$\mathscr{B}$ (\textbf{Bayes Factor}) \\ \hline
\multicolumn{7}{l}{\textbf{Approach 1}}                                                                             \\ \hline
Baseline           &  1.362 $\pm 0.001$  & 1663 $^{+18}_{-16}$ & -2.559 $\pm 0.08$ & -3.66 $\pm 0.11$ & -3.37 $\pm 0.08$ & 890.4 $\pm$ 0.2                \\
ASteRA             & 1.391 $\pm 0.003$  & 1452 $^{+24}_{-20}$ & -3.03 $\pm 0.012$ & -4.05 $\pm 0.13$  &  -4.04 $^{+0.13}_{-0.12}$ & 921.9 $\pm$ 0.2 \textbf{(31.5)}               \\ \hline
\multicolumn{7}{l}{\textbf{Approach 2}}                                                                             \\ \hline
Baseline (1 $\mu$m)  & 1.374 $\pm 0.08$ & 1600 $^{+29}_{-25}$ & -3.01 $^{+0.39}_{-0.47}$ & -4.08 $^{+0.39}_{-0.47}$  & -3.86 $^{+0.40}_{-0.47}$  & 871.6 $\pm$ 0.1                \\
ASteRA (1 $\mu$m)    & 1.392 $^{+0.010}_{-0.008}$ & 1455 $^{+27}_{-24}$  & -3.13 $^{+0.48}_{-0.55}$  & 4.16 $^{+0.50}_{-0.54}$  & -4.13 $^{+0.48}_{-0.55}$ & 896.1 $\pm$ 0.2 \textbf{(24.5)}               \\
Baseline (1.5 $\mu$m) & 1.392 $^{+0.009}_{-0.011}$ & 1469 $^{+27}_{-26}$  & -3.48 $^{+0.62}_{-0.53}$  & -4.50 $^{+0.62}_{-0.53}$  & -4.45 $^{+0.63}_{-0.52}$ & 839.9 $\pm$ 0.1                \\
ASteRA (1.5 $\mu$m)  & 1.393 $\pm 0.010$ & 1460 $^{+28}_{-24}$ & -3.40 $^{+0.62}_{-0.57}$ & -4.43 $^{+0.63}_{-0.56}$ & -4.39 $^{+0.62}_{-0.56}$ & 840.2 $\pm$ 0.1 \textbf{(0.3)}               \\
Baseline (2 $\mu$m)  &  1.392 $^{+0.010}_{-0.011}$ & 1458 $^{+31}_{-29}$  & -3.44 $^{+0.61}_{-0.55}$  & -4.48 $^{+0.61}_{-0.53}$ & -4.44 $^{+0.62}_{-0.57}$ & 754.5 $\pm$ 0.1                \\
ASteRA (2 $\mu$m)   & 1.395 $^{+0.010}_{-0.012}$  & 1455 $^{+32}_{-27}$ & -3.45 $^{+0.69}_{-0.57}$ & -4.47 $^{+0.68}_{-0.57}$ & -4.45 $^{+0.69}_{-0.56}$ &  754.6 $\pm$ 0.1 \textbf{(0.1)}         \\ \hline
\multicolumn{7}{l}{\textbf{Approach 3}}                                                                             \\ \hline
Baseline    & 1.403 $\pm 0.001$ & 1394 $\pm 17$  & -3.28 $\pm 0.07$  & -4.30 $^{+0.10}_{-0.09}$ & -4.38 $\pm 0.07$ & 918.6 $\pm$ 0.2                \\
ASteRA     & 1.394 $\pm 0.003$ & 1442 $^{+20}_{-18}$  & -3.13 $^{+0.09}_{-0.10}$  & -4.16 $\pm 0.11$ & -4.16 $\pm 0.11$ & 922.0 $\pm$ 0.2 \textbf{(3.4)}               \\ \hline
\end{tabular}
\end{table*}

\begin{table*}
\label{tab:spot_fac_results}
\caption{Retrieval results for the three approaches applied to the mixed spot- and facula-contaminated stacked spectrum. ln$\mathscr{B}$ is the Bayesian evidence returned and the Bayes factors comparing baseline and ASteRA model pairs are given next to the preferred model.}
\centering
\begin{tabular}{lllllll}
\hline \hline
                   & $R_{\rm p}$ [$R_{\rm Jup}$] & $T_{\rm p}$ [K] & log(H$_2$O) & log(CO$_2$) & log(CH$_4$) & ln$\mathscr{B}$ (\textbf{Bayes Factor}) \\ \hline
\multicolumn{7}{l}{\textbf{Approach 1}}                                                                             \\ \hline
Baseline           &  1.456 $\pm 0.006$  & 1036 $^{+15}_{-13}$ & -3.83 $\pm 0.07$  & -4.94 $\pm 0.08$  & -5.61 $\pm 0.06$  & 452.1 $\pm$ 0.2                \\
ASteRA             & 1.390 $\pm 0.003$ & 1470 $^{+24}_{-21}$& -3.20 $^{+0.17}_{-0.16}$ & -4.20 $\pm 0.17$ & -4.20 $^{+0.17}_{-0.16}$  & 915.2 $\pm$ 0.2 \textbf{(463.1)}               \\ \hline
\multicolumn{7}{l}{\textbf{Approach 2}}                                                                             \\ \hline
Baseline (1 $\mu$m)  & 1.439 $\pm 0.003$ & 1085 $^{+13}_{-16}$ & -2.74 $\pm 0.21$  & -4.02 $^{+0.21}_{-0.20}$ & -4.60 $\pm 0.20$ & 505.9 $\pm$0.2               \\
ASteRA (1 $\mu$m)    & 1.392 $\pm 0.010$  & 1470 $^{+30}_{-25}$ & -3.26 $\pm 0.56$ & -4.26 $^{+0.56}_{-0.55}$ & -4.27 $^{+0.57}_{-0.55}$ & 889.7 $\pm$ 0.2 \textbf{(383.8)}               \\
Baseline (1.5 $\mu$m) & 1.409 $^{+0.005}_{-0.004}$ & 1385 $^{+23}_{-26}$ & -2.56 $^{+0.25}_{-0.27}$ & -3.63 $\pm 0.27$ & -3.80 $\pm 0.27$ & 734.0 $\pm$ 0.1                \\
ASteRA (1.5 $\mu$m)  & 1.393 $^{+0.010}_{-0.011}$  & 1472 $^{+30}_{-26}$ & -3.36 $^{+0.62}_{-0.57}$ & -4.36 $^{+0.62}_{-0.55}$ & -4.37 $^{+0.64}_{-0.54}$ & 833.3 $\pm$ 0.2 \textbf{(99.3)}               \\
Baseline (2 $\mu$m)  & 1.416 $\pm 0.006$  & 1370 $^{+29}_{-31}$  & -2.96 $^{+0.33}_{-0.38}$ & -3.99 $^{+0.33}_{-0.37}$  & -4.18 $^{+0.33}_{-0.36}$  & 674.1 $\pm$ 0.1                \\
ASteRA (2 $\mu$m)    & 1.392 $\pm 0.010$  & 1468 $^{+31}_{-28}$  & -3.43 $^{+0.57}_{-0.50}$  & -4.47 $^{+0.57}_{-0.50}$  & -4.46 $^{+0.57}_{-0.50}$ & 750.0 $\pm$ 0.2 \textbf{(75.9)}               \\ \hline
\multicolumn{7}{l}{\textbf{Approach 3}}                                                                             \\ \hline
Baseline           & 1.393 $\pm 0.001$ & 1452 $^{+14}_{-16}$ & -3.03 $\pm 0.07$ & -4.03 $\pm 0.09$ & -4.06 $^{+0.07}_{-0.06}$ & 919.2 $\pm$ 0.2 \textbf{(0.2)}               \\
ASteRA             & 1.393 $^{+0.002}_{-0.003}$ & 1458 $^{+20}_{-17}$ & -3.06 $^{+0.09}_{-0.11}$ & -4.06 $^{+0.11}_{-0.12}$ & -4.09 $^{+0.10}_{-0.11}$ & 919.0 $\pm$ 0.2                \\ \hline
\end{tabular}
\end{table*}

\begin{figure*}
    \centering
    \includegraphics[width=1.8\columnwidth]{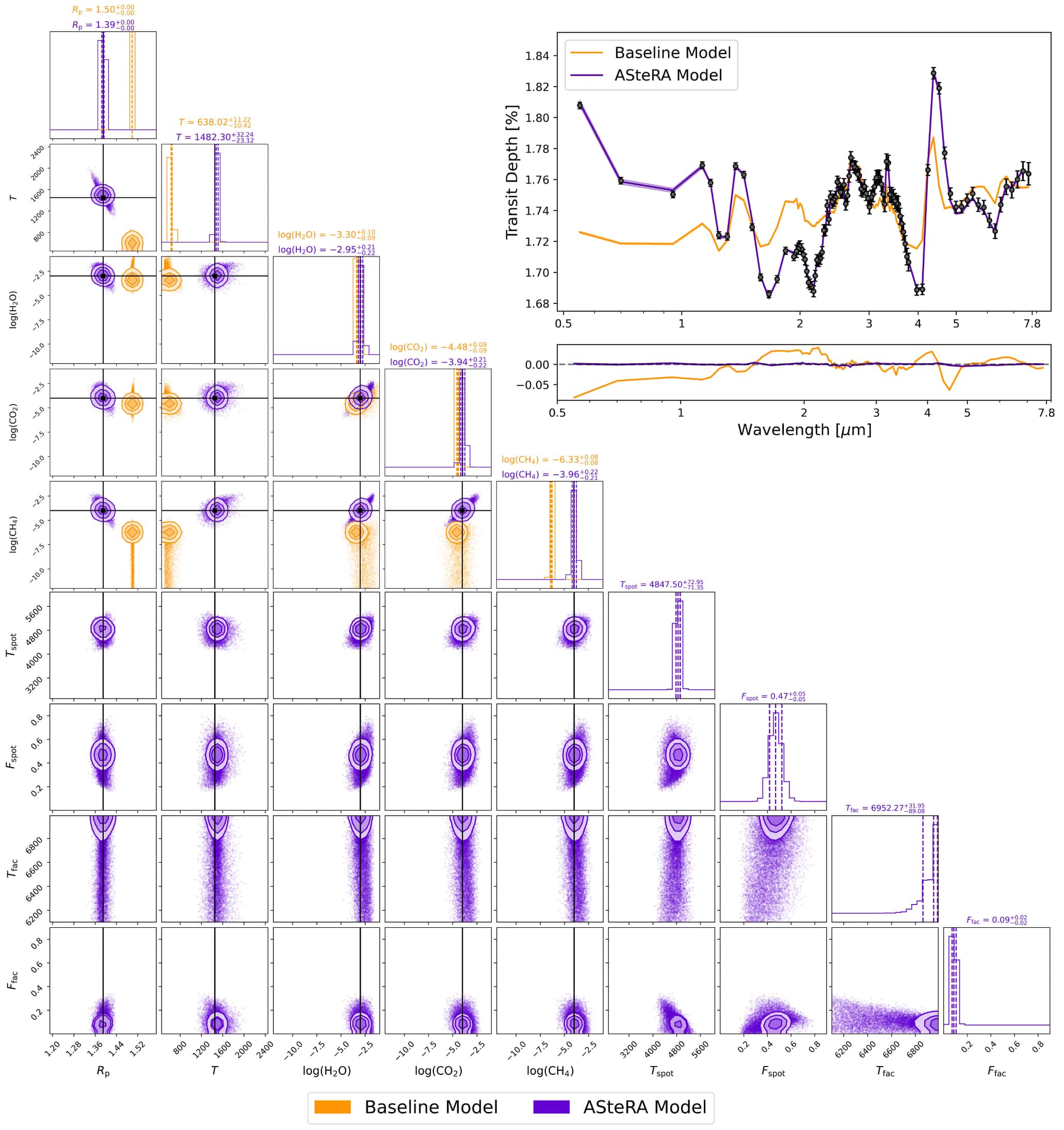}
    \caption{Overplotted posterior distributions showing the results of the baseline (\textit{orange}) and ASteRA (\textit{purple}) retrievals for Approach 1 when the four most spot-contaminated visits are stacked. \textit{Inset} - The best-fit spectra retrieved by each model and their associated residuals.}
    \label{fig:spot_uncorr_posterior}
\end{figure*}

\begin{figure*}
    \centering
    \includegraphics[width=1.8\columnwidth]{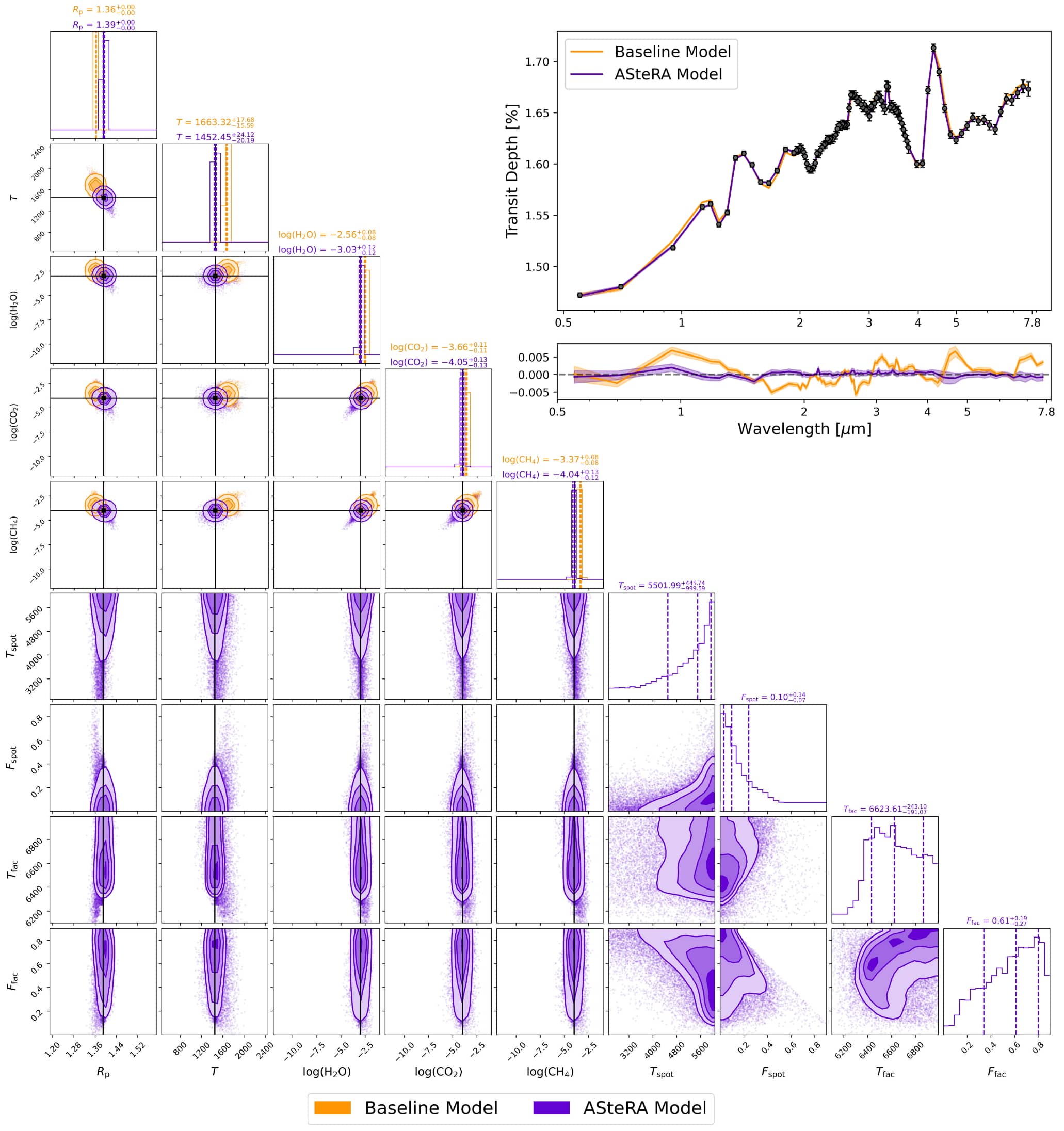}
    \caption{As for Figure~\ref{fig:spot_uncorr_posterior} but for the combination of the four most facula-contaminated visits.}
    \label{fig:fac_uncorr_posterior}
\end{figure*}

\begin{figure*}
    \centering
    \includegraphics[width=1.8\columnwidth]{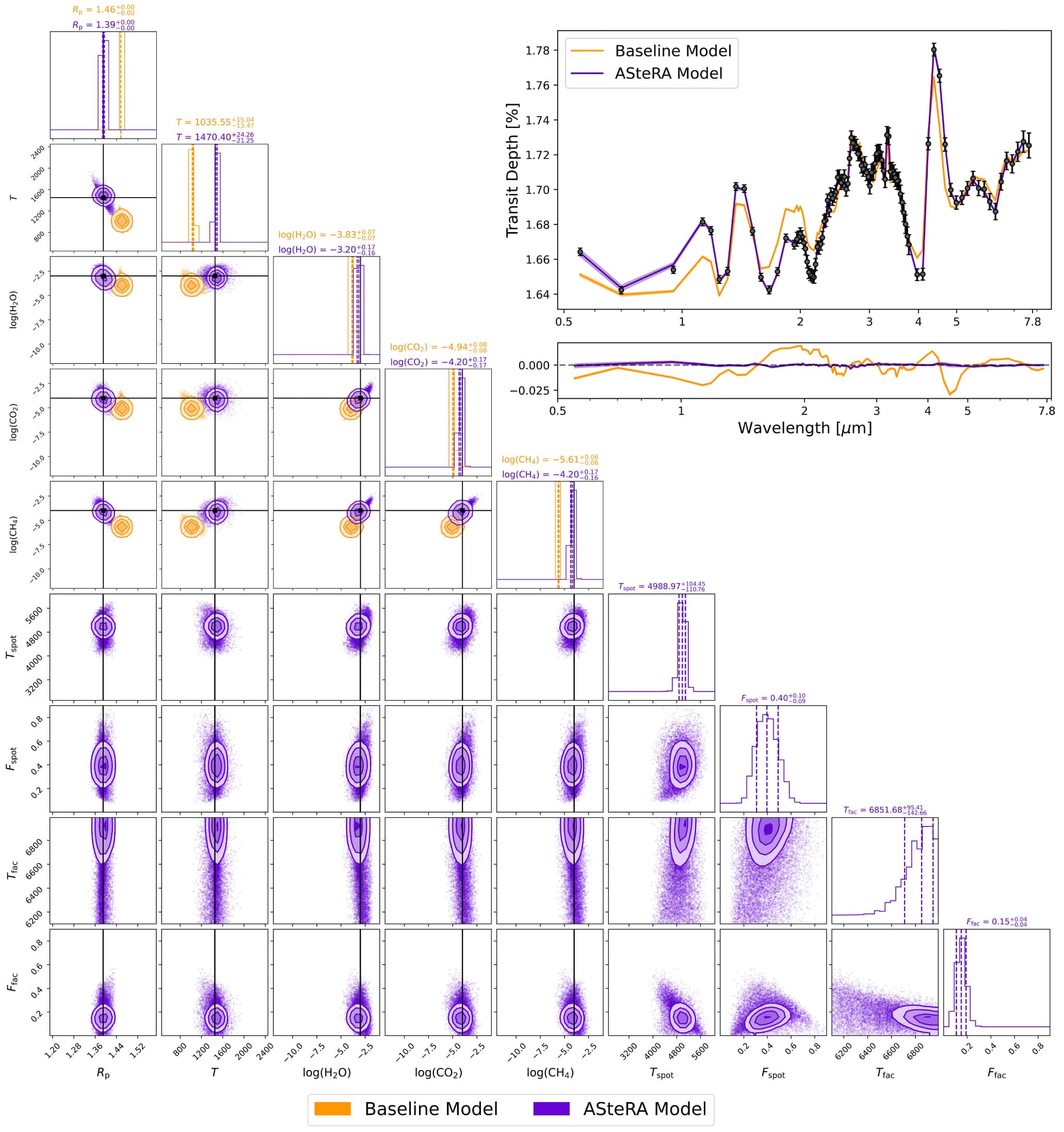}
    \caption{As for Figure~\ref{fig:spot_uncorr_posterior} but for the combination of the two most spot-contaminated and two most facula-contaminated visits.}
    \label{fig:spot_fac_uncorr_posterior}
\end{figure*}

\begin{figure*}
    \centering
    \includegraphics[width=1.8\columnwidth]{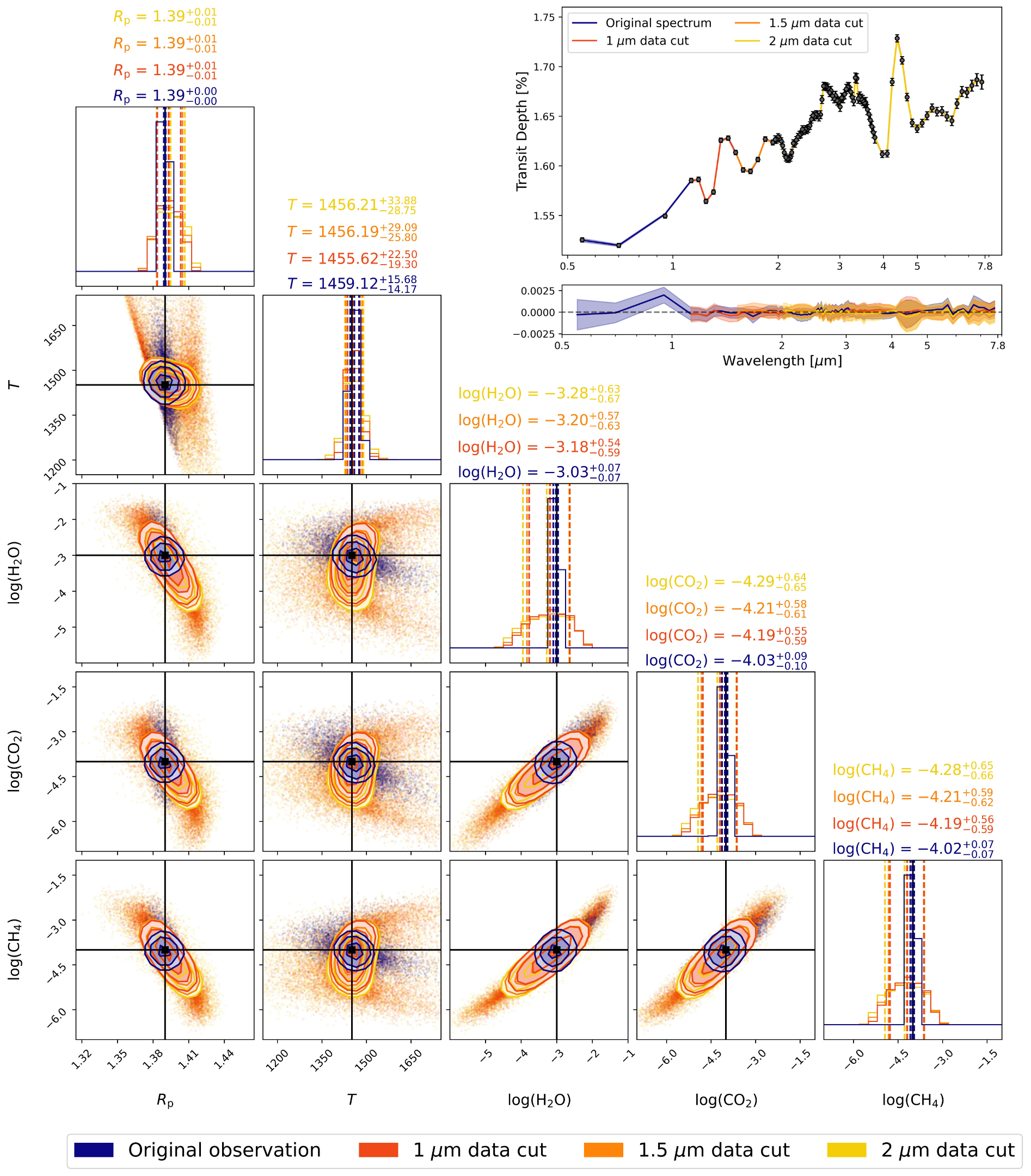}
    \caption{Posterior distributions for the data cut method of Approach 2 applied to the uncontaminated spectrum. Figure elements are the same as for Fig.~\ref{fig:spot_base_datacut_posterior}.}
    \label{fig:uncont_datacut_posterior}
\end{figure*}

\begin{figure*}
    \centering
    \includegraphics[width=1.8\columnwidth]{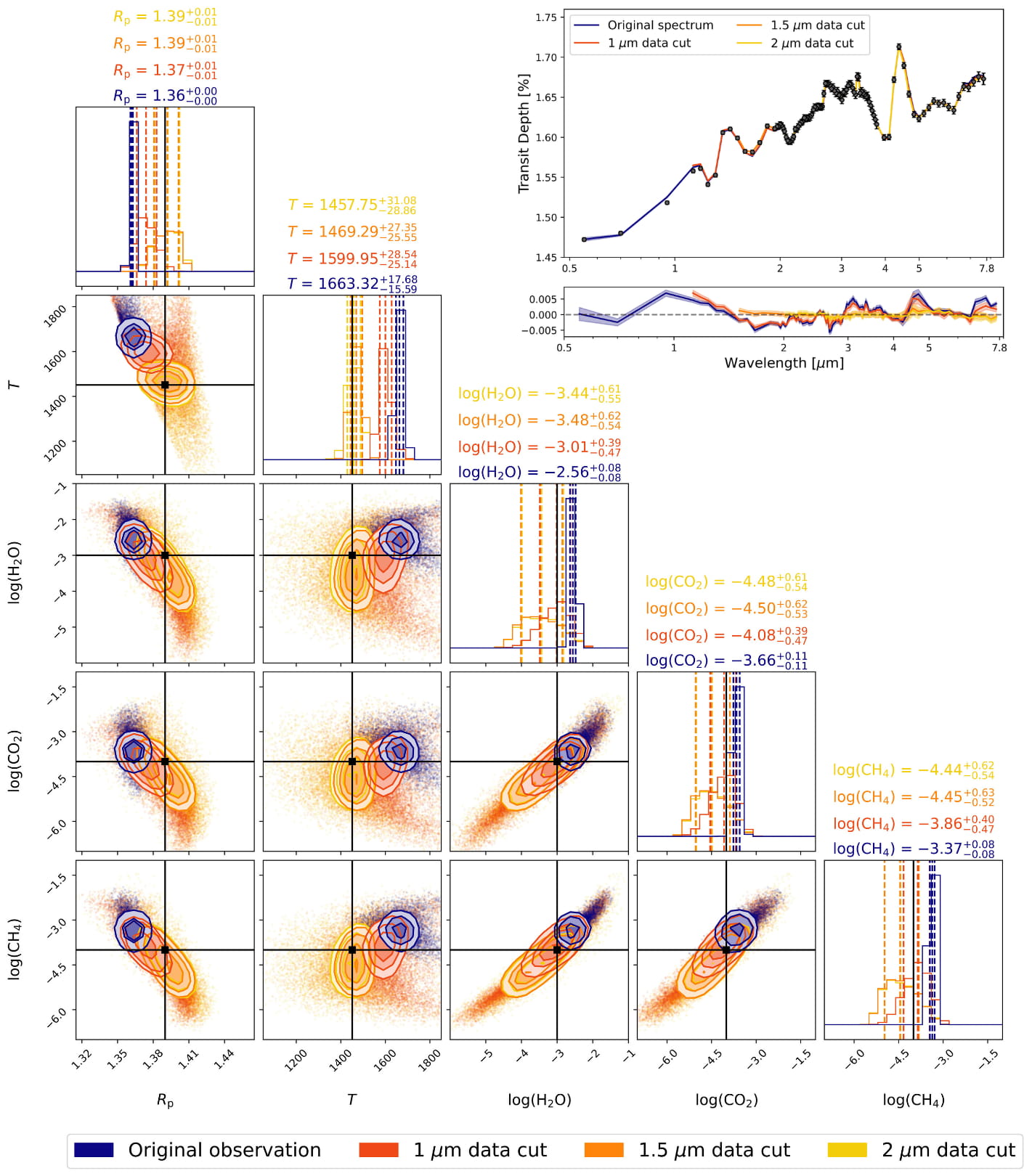}
    \caption{Posterior distributions for the data cut method of Approach 2 applied to the stacked spectrum consisting of the four most facula-contaminated visits. Figure elements are the same as for Fig.~\ref{fig:spot_base_datacut_posterior}.}
    \label{fig:fac_datacut_posterior}
\end{figure*}

\begin{figure*}
    \centering
    \includegraphics[width=1.8\columnwidth]{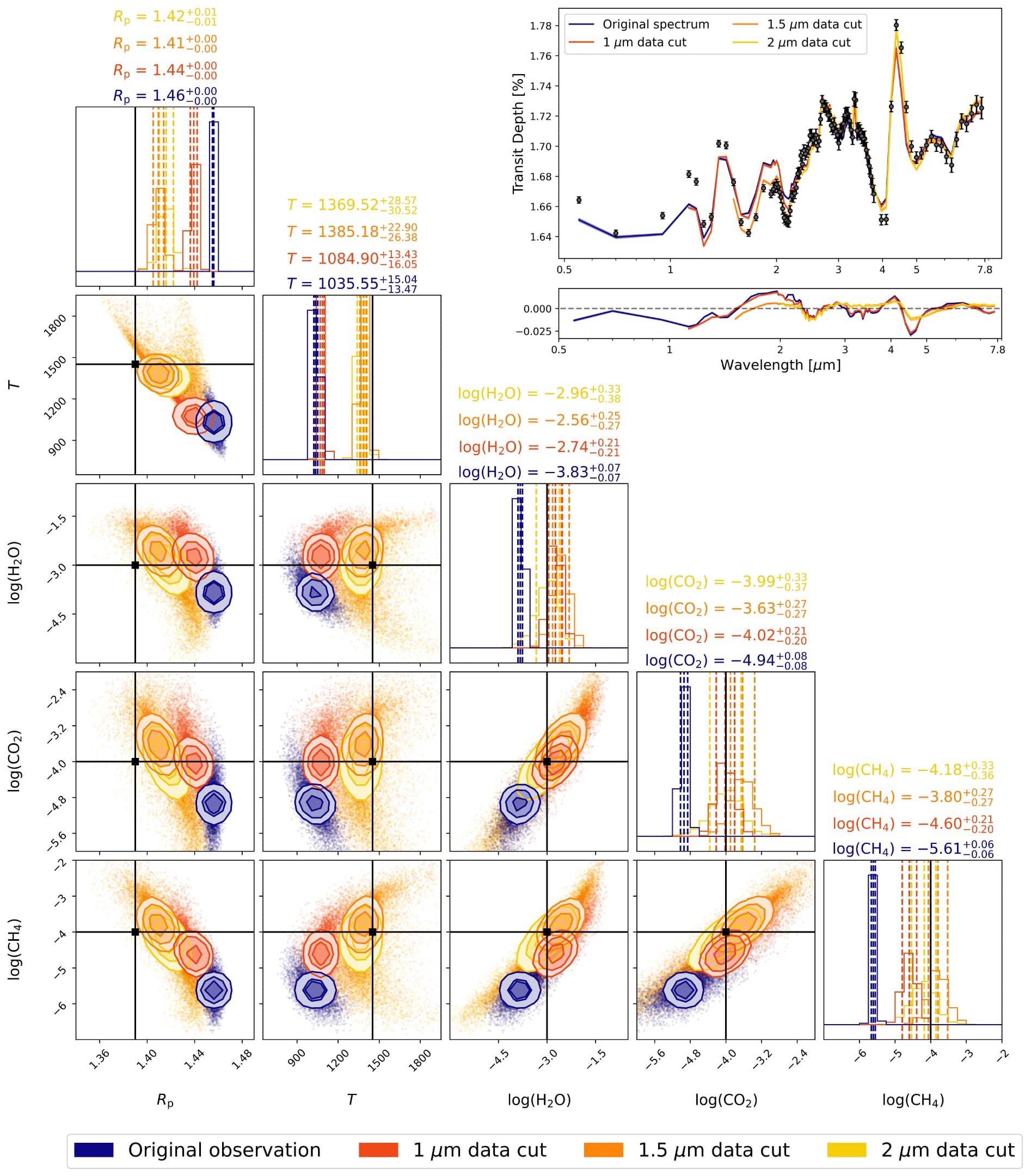}
    \caption{Posterior distributions for the data cut method of Approach 2 applied to the stacked spectrum consisting of the two most spot- and facula contaminated visits. Figure elements are the same as for Fig.~\ref{fig:spot_base_datacut_posterior}.}
    \label{fig:spot_fac_datacut_posterior}
\end{figure*}

\begin{figure*}
    \centering
    \includegraphics[width=1.5\columnwidth]{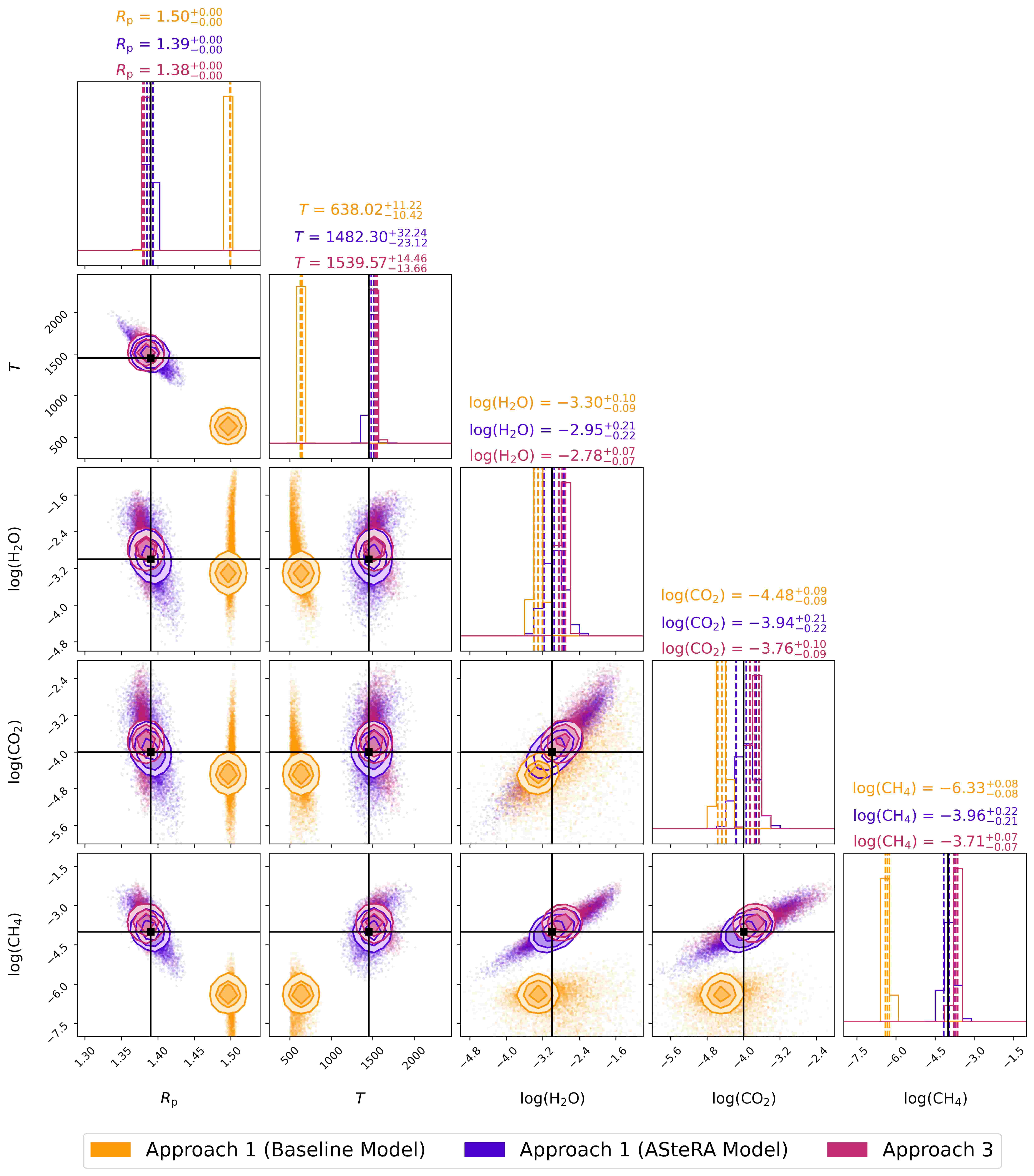}
    \caption{Posterior distributions of the planetary parameters retrieved with Approach 3 (\textit{pink}) compared to the baseline (\textit{orange}) and ASteRA (\textit{purple}) retrievals conducted under Approach 1 for the spot-contaminated stacked spectrum.}
    \label{fig:spot_approach3}
\end{figure*}

\begin{figure*}
    \centering
    \includegraphics[width=1.5\columnwidth]{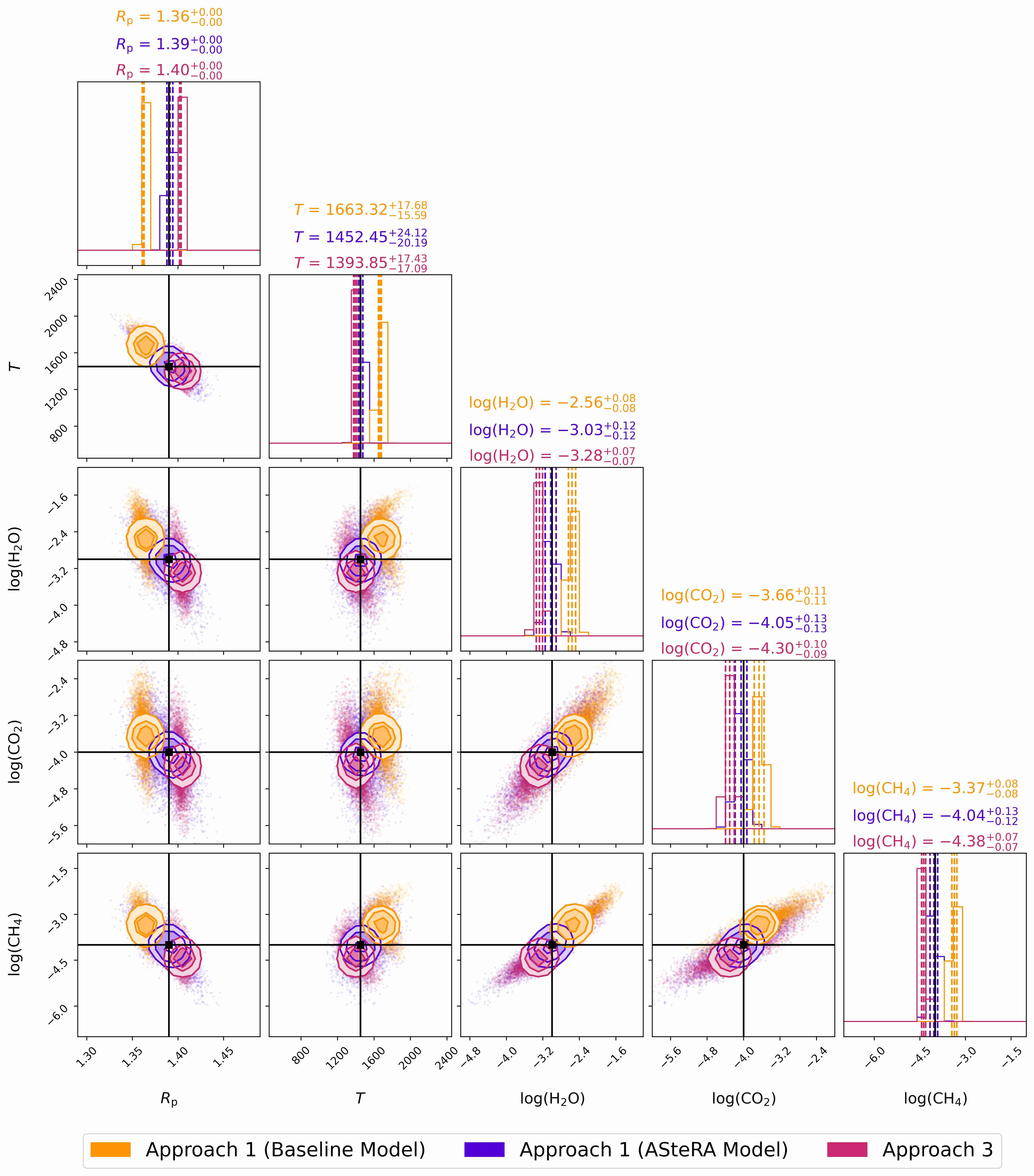}
    \caption{As for Fig~\ref{fig:spot_approach3} but for the facula-contaminated stacked spectrum.}
    \label{fig:fac_approach3}
\end{figure*}

\begin{figure*}
    \centering
    \includegraphics[width=1.5\columnwidth]{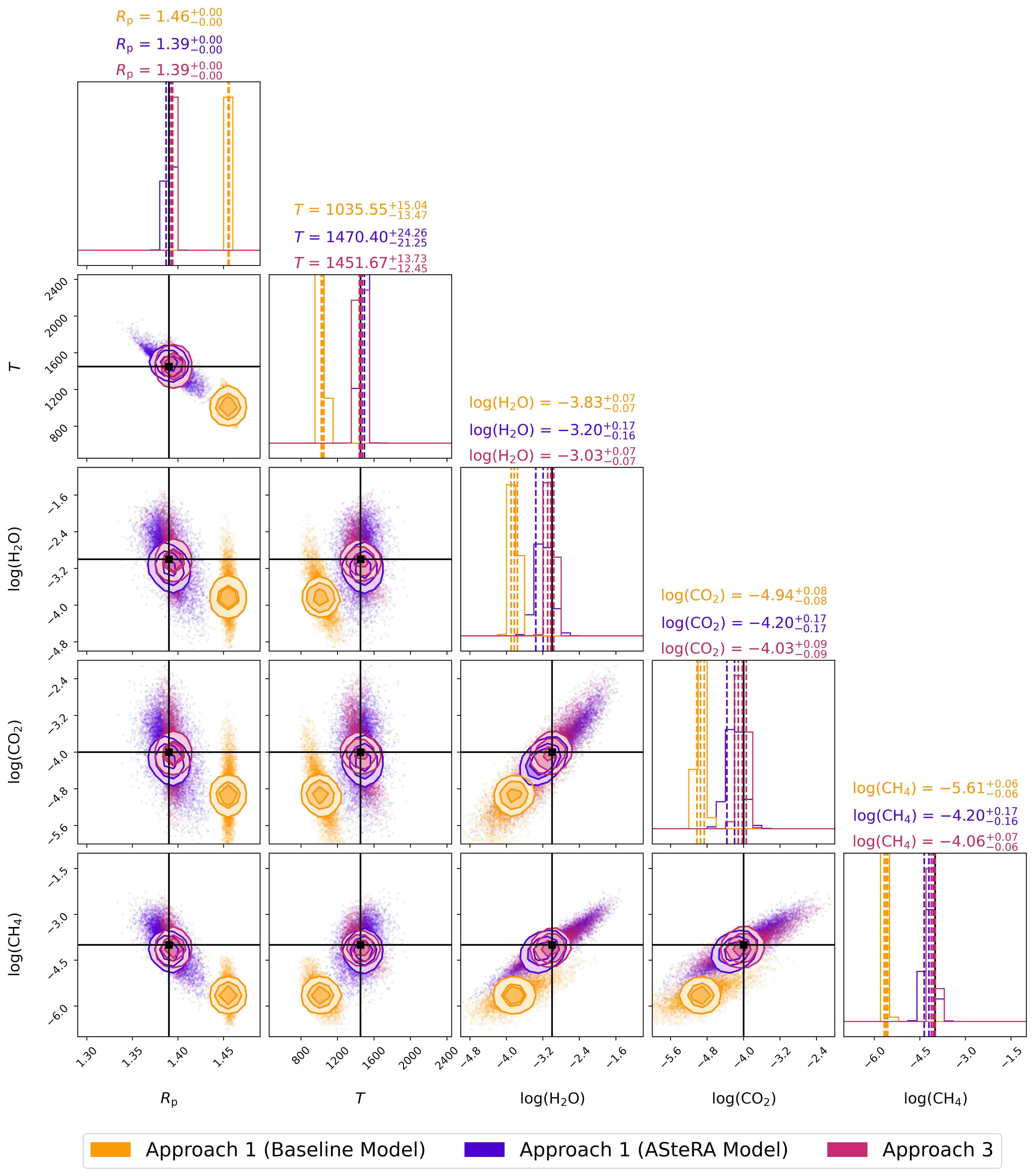}
    \caption{As for Fig~\ref{fig:spot_approach3} but for the stacked spectrum that combines both spot- and facula-contaminated visits.}
    \label{fig:spot_fac_approach3}
\end{figure*}


\bsp	
\label{lastpage}
\end{document}